\documentclass[11pt,oneside,a4paper,onecolumn]{elsarticle}

\usepackage{times}
\usepackage{fullpage}

\usepackage{gensymb}
\usepackage{alltt}
\usepackage{upquote}
\usepackage{float}
\usepackage{array}
\usepackage{pgfplots}
\usetikzlibrary{positioning}
\usepackage{graphicx}
\usepackage[toc,page]{appendix}
\usepackage[utf8]{inputenc}
\usepackage{amsmath,mathtools}
\usepackage{amsthm}
\usepackage{amsfonts}
\usepackage{amssymb}
\usepackage{enumitem}
\usepackage{caption}
\usepackage{subcaption}
\usepackage{xcolor}
\usepackage{scalerel}
\usepackage{algorithm}
\usepackage{algorithmic}
\newtheorem{rem}{Remark}
\usepackage{multicol}
\usepackage{multirow}
\usepackage{url}
\usepackage{booktabs}
\usepackage{nicefrac}

\usepackage{soul}
\usepackage{slashed}
\usetikzlibrary{patterns}
\usetikzlibrary{shapes.arrows}
\usetikzlibrary{fillbetween,backgrounds}
\usepackage{placeins}
\pgfplotsset{compat=newest}
\usepackage{diagbox}
\usetikzlibrary{arrows.meta}
\pgfdeclarelayer{nodelayer}
\pgfdeclarelayer{edgelayer}
\pgfsetlayers{background,nodelayer,edgelayer,main}

\usepackage[final]{hyperref} % adds hyper links inside the generated pdf file
\hypersetup{
	colorlinks=true,       % false: boxed links; true: colored links
	linkcolor=blue,        % color of internal links
	citecolor=blue,        % color of links to bibliography
	filecolor=magenta,     % color of file links
	urlcolor=blue         
}

\newcolumntype{M}[1]{>{\centering\arraybackslash}m{#1}}
\newcolumntype{N}{@{}m{0pt}@{}}

\tikzset{>=latex}
\def\@author#1{\g@addto@macro\elsauthors{\normalsize%
    \def\baselinestretch{1}%
    \upshape\authorsep#1\unskip\textsuperscript{%
      \ifx\@fnmark\@empty\else\unskip\sep\@fnmark\let\sep=,\fi
      \ifx\@corref\@empty\else\unskip\sep\@corref\let\sep=,\fi
      }%
    \def\authorsep{\unskip,\space}%
    \global\let\@fnmark\@empty
    \global\let\@corref\@empty
    \global\let\sep\@empty}%
    \@eadauthor={#1}
}

\graphicspath{{Images/}}

\begin{document}
\begin{frontmatter}
\title{A Finite Element-Inspired Hypergraph Neural Network:\\ Application to Fluid Dynamics Simulations}
\author[ubc]{Rui Gao}
\ead{garrygao@mail.ubc.ca}

\author[ubc]{Indu Kant Deo}
\ead{indukant@mail.ubc.ca}

\author[ubc]{Rajeev K. Jaiman\corref{cor1}}
\ead{rjaiman@mail.ubc.ca}
\cortext[cor1]{Corresponding author}
\address[ubc]{Department of Mechanical Engineering, The University of British Columbia, Vancouver, BC V6T 1Z4}

\begin{abstract}
An emerging trend in deep learning research focuses on the applications of graph neural networks (GNNs) for mesh-based continuum mechanics simulations. Most of these learning frameworks operate on graphs wherein each edge connects two nodes. Inspired by the data connectivity in the finite element method, we present a method to construct a hypergraph by connecting the nodes by elements rather than edges. A hypergraph message-passing network is defined on such a node-element hypergraph that mimics the calculation process of local stiffness matrices. We term this method a finite element-inspired hypergraph neural network, in short FEIH($\phi$)-GNN. We further equip the proposed network with rotation equivariance, and explore its capability for modeling unsteady fluid flow systems. The effectiveness of the network is demonstrated on two common benchmark problems, namely the fluid flow around a circular cylinder and airfoil configurations. Stabilized and accurate temporal roll-out predictions can be obtained using the $\phi$-GNN framework within the interpolation Reynolds number range. The network is also able to extrapolate moderately towards higher Reynolds number domain out of the training range.
\smallskip
\smallskip

 \end{abstract}
\end{frontmatter}

%\begin{linenumbers}

\section{Introduction}
\label{sec:intro}
There has been an increasing interest in predicting and controlling the spatial-temporal dynamics of fluid flow based on the solution of Navier-Stokes equations \cite{collis2004issues,joslin2009fundamentals}. Since analytical solutions are usually not available, numerical solutions on discretized space and time domains are considered for predictive modeling. Leveraging state-of-the-art computational fluid dynamics (CFD) approaches based on finite volume \cite{leveque2002finite} or finite element \cite{hughes2012finite,johnson2012numerical} methods, one could obtain high-fidelity solutions that can be suitable for downstream design optimization and control purposes. However, the cost of performing such simulations is significant, and becomes prohibitively high for complex problems arising from real-world applications.

This limitation of traditional CFD approaches has inspired the development of data-driven projection-based reduced-order modeling techniques. Such models are usually used in an offline-online manner. In the offline stage, an approximation of the governing flow dynamics in a low-order linear subspace is constructed based on available fluid flow data collected. This approximation reduces the complexity of the problem in the online stage, making it possible to acquire fast, accurate predictions. Popular methods in this category include proper orthogonal decomposition (POD) \cite{Lumley1967,Sirovich1987}, dynamic mode decomposition (DMD) \cite{Schmid2010}, along with many variants (e.g., \cite{Towne2018,Schmidt2019,Zhang2019}). However, these methods encounter difficulty when applied to scenarios with high Reynolds numbers and convection-dominated problems, whereas one needs a significantly large number of linear subspaces to achieve a satisfactory approximation. 

In the last decade, deep neural networks have been explored as alternatives to the aforementioned techniques. The convolutional neural network, especially the convolutional encoder-propagator-decoder architecture, has been adopted in many applications, including flow around rigid bodies in both 2D \cite{Gonzalez2018,Thuerey2020,Bukka2021,Pant2021} and 3D \cite{Gupta2022a} scenarios, interactions between fluid flow and moving solid bodies \cite{Gupta2022b,Zhang2022,fan2023differentiable}, wave and sound propagations \cite{deo2022predicting,Mallik2022}, and many more. By exploiting the computational power of modern graphical processing units (GPU), these convolutional neural networks are usually orders of magnitude faster than the traditional full-order CFD simulations.

While convolutional neural networks have achieved numerous successes, they are inherently restricted to be performed on a uniform Cartesian grid, which limits their capabilities in scenarios with complex geometries. Furthermore, due to the same reason, one cannot have different grid resolutions between the area of minor interest and the area with important dynamics. Under such restriction, one have to either introduce a dense Cartesian grid covering the whole field, or tolerate low resolution at regions of interest. The former is not always feasible due to computational resource limitations and prediction speed requirements. Therefore, the latter choice is usually inevitable, leading to the loss of fine physical details and subsequent difficulties in calculating important physical statistics such as the lift and drag forces.

\begin{figure}[t]
	\centering
	\includegraphics[]{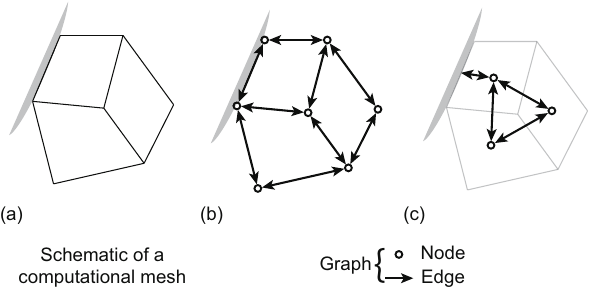}
	\caption{Conversion of a representative computational mesh to graph: (a) schematic representing a portion of computational mesh, (b)-(c) two possible approaches to convert the mesh to a graph.}
	\label{fig:mesh2negraph}
\end{figure}

It turns out that the (unstructured) computational mesh used for traditional CFD simulations can be converted to a graph $\mathcal{G}=(\mathcal{V},\mathcal{E})$ with nodes $\mathcal{V}$ connected by edges $\mathcal{E}$ rather easily. Graph neural networks that works on such graphs therefore have the ability to address the aforementioned difficulty with convolutional neural networks. One can consider two approaches for the conversion of a mesh (Fig.~\ref{fig:mesh2negraph}a) into a graph. One approach (Fig.~\ref{fig:mesh2negraph}b) converts each vertex of the mesh to a node, and each cell boundary between two vertices to two directed edges. The other approach (Fig.~\ref{fig:mesh2negraph}c) converts each cell within the mesh into a node, while each border between two neighboring cells is converted to two directed edges.

With the system states originally attached to the mesh converted to the node and edge features, a graph neural network can be applied to learn the temporal evolution of these features, effectively serving as a surrogate to the traditional CFD model of the system. As a result, GNNs can maintain a significantly better resolution for important regions within the domain, while keeping the same mesh size as the CNNs. With this clear advantage over CNN, graph neural networks are recently introduced to the field of continuum mechanics \cite{belbute2020combining,pfaff2020learning}, inspiring a surge of works in the past three years. In particular, the encode-process-decode architecture with \emph{generalized graph message-passing} layers \cite{pfaff2020learning,Battaglia2018,sanchez2018graph} see popularity in many applications, including flow around fixed bodies such as cylinders, airfoils or buildings \cite{pfaff2020learning,Lino2022,shao2023pignn}, reacting flows \cite{Xu2021}, flow field super-resolution \cite{belbute2020combining, Xu2021}, flow field completion \cite{he2022flow}, etc. Additional techniques and designs are also being developed to work with this architecture, like on-the-fly graph adaptation \cite{pfaff2020learning}, multi-grid methods \cite{Lino2022,yang2022amgnet,fortunato2022multiscale,cao2022bi}, rotation equivariance \cite{Lino2022}, among many more.

For a graph neural network that fits into the generalized graph message-passing \cite{Battaglia2018,sanchez2018graph} framework, each message-passing layer can be written as the combination of an edge update stage
\begin{subequations}
	\label{eq:nemp}
	\begin{equation}
		\label{eq:edgene}
		e_{ij}^\prime=\phi^e(e_{ij},v_i,v_j),
	\end{equation}
	and a node update stage,
	\begin{equation}
		\label{eq:nodene}
		v_i^\prime=\phi^v(v_i,\operatorname{AGG}_j{e_{ji}^\prime}),
	\end{equation}
\end{subequations}
in which $v_i$ and $v_j$ denote the node features attached to the nodes $i$ and $j$ respectively, $e_{ij}$ denotes the edge feature corresponding to the directed edge pointing from node $i$ to node $j$, and the superscript $(\cdot)^\prime$ denotes the updated features. The edge update function $\phi^e$ and the node update function $\phi^v$ are some nonlinear functions, e.g., multi-layer perceptrons. The function $\operatorname{AGG}$ is a permutation-invariant aggregation function that aggregates the information from all the edges pointing to each node $i$.

From a computational mechanics point of view, such a message-passing network defined on the graph $\mathcal{G}$ mimics the finite volume method \cite{leveque2002finite}. Interpreting each node as a control volume, and each edge as the boundary between neighboring control volumes, the edge update function $\phi^e$ calculates the flux of information between the two control volumes, and the node update function $\phi^v$ evaluates the information in the control volume based on the aggregated incoming flux. With a residual link \cite{he2016} for node and edge features, each layer or step within the message-passing network can be explained as one iteration in the iterative approximation towards the ground truth flux and cell information updates.

Besides the finite volume method, many other approaches are available in computational mechanics. The finite element method, for example, has been used as inspiration for some recent works. Alet et al. \cite{alet2019graph} constructs encoders and decoders that interpolate between data at random points within the field and data on the nodes of the graph, mimicking the behavior of shape function-based interpolation within the element in finite element methods. Gao et al. \cite{Gao2022}, more recently, proposed to calculate the loss by integrating the prediction error over the simulation domain using Gaussian quadrature integration with high-order shape functions instead of the traditional mean-squared loss on nodes. 

\begin{figure}[t]
	\centering
	\includegraphics[]{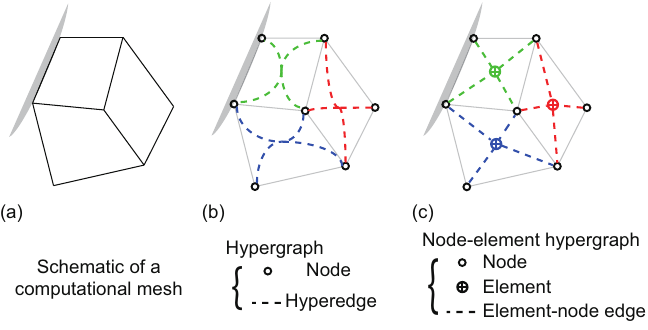}
	\caption{Conversion of a computational mesh to hypergraph: (a) schematic representing a portion of computational mesh, (b) converted hypergraph, (c) converted node-element hypergraph.}
	\label{fig:mesh2hypergraph}
\end{figure}

In contrast to these works, we take inspiration from the data connectivity of the finite element method and connect the nodes $\mathcal{V}$ by elements $\mathcal{E}_\square$, effectively forming a hypergraph, illustrated in Fig.~\ref{fig:mesh2hypergraph}b. Further modifications of the hypergraph as detailed in Sec. \ref{sec:mesh2ne+graph} lead to a node-element hypergraph (Fig.~\ref{fig:mesh2hypergraph}c). We further mimic the local stiffness matrix calculation process in the finite element analysis, and design a hypergraph message-passing network that works with the node-element hypergraph converted from a computational mesh.  We refer to this as a finite element-inspired hypergraph neural network, or in short, FEIH($\phi$)-GNN).  We further equip the $\phi$-GNN with rotational equivariance via appropriate input and output feature transformations and apply the network for the simulation of fluid flow around obstacles. For the assessment of our $\phi$-GNN, numerical experiments are systematically performed on two classic fluid flow configurations, namely the flow around a cylinder and the flow around an airfoil. Through our numerical experiments, we demonstrate stabilized and accurate predictions of the flow field as well as the lift and drag statistics calculated from it.

To the best of our knowledge, the most closely-related work to this research is the recent work by Lienen and G\"{u}nnemann \cite{lienen2022learning}, who employed a single hypergraph message-passing step to estimate the time derivatives of system states at each node, which are then sent to an ODE solver to generate predictions of the system states in the future time steps. In contrast to the approach adopted in \cite{lienen2022learning}, our work has similarities with that of Pfaff et al. \cite{pfaff2020learning}. For example, we introduce more non-local information by stacking multiple hypergraph message-passing layers, and we directly predict the difference of system states between neighboring time steps (i.e., using a forward Euler time discretization) rather than relying on an ODE solver. We also target stable and accurate predictions over a long horizon, with tests on four thousand time steps into the future starting from one single time step, while the results reported by Lienen and G\"{u}nnemann only include predictions within a short period of future (60-time steps at most).

The remaining part of this paper is organized as follows: Section \ref{sec:method} presents the conversion from mesh to node-element hypergraph, the derivation of the $\phi$-GNN layers from the stiffness matrix assembly process, as well as the network architecture. Section \ref{sec:model} discusses the application of $\phi$-GNN in fluid flow simulations. Sec. \ref{sec:exp} discusses the experimental setup and model training details, with the results presented in Sec. \ref{sec:results}. We conclude the work in Sec. \ref{sec:summary}.

\section{Methodology}
\label{sec:method}
In this section, we introduce our node-element hypergraph together with the hypergraph message-passing layer that works with it. Starting from a general continuum dynamical system, we discretize it in space and time, and then describe the target function to be approximated with our $\phi$-GNN model in Sec. \ref{sec:rom}. We proceed to explain the conversion of the mesh employed to discretize the system in space to a node-element hypergraph in Sec. \ref{sec:mesh2ne+graph}, followed by a detailed derivation of the hypergraph message-passing layer from the local stiffness matrix calculation process in Sec. \ref{sec:elemmp} and \ref{sec:mlp}. The complete architecture of the network is presented in Sec. \ref{sec:network}.

\subsection{Surrogate model of a continuum system}
\label{sec:rom}
Consider a continuum dynamical system defined in a bounded spatial domain $\Omega$, with governing equation in a general form
\begin{equation}
\frac{\partial q(\boldsymbol{x},t)}{\partial t}=G(q(\boldsymbol{x},t),F(\boldsymbol{x},t)),
\end{equation}
in which $q$ is the system state and $F$ denotes any input to the system. We discretize in time with the forward Euler scheme
\begin{equation}
q(\boldsymbol{x},t_{n+1})-q(\boldsymbol{x},t_n)=\tilde{G}(q(\boldsymbol{x},t_n),F(\boldsymbol{x},t_n)),
\end{equation}
where $t_n$ denotes the $n$-th time step. Further discretization in space with a mesh leads to
\begin{equation}
\label{eq:rom}
\boldsymbol{Q}(t_{n+1})-\boldsymbol{Q}(t_n)=\tilde{\boldsymbol{G}}(\boldsymbol{q}(t_n),\boldsymbol{F}(t_n))\approx\widehat{\boldsymbol{G}}(\boldsymbol{q}(t_n),\boldsymbol{F}(t_n)).
\end{equation}
in which $\boldsymbol{Q}(t_n)$ denotes the system state vector at time step $t_n$, and $\boldsymbol{F}(t_n)$ denotes the system input vector at time step $t_n$. The function $\tilde{\boldsymbol{G}}$ that governs the evolution of the system in time can be approximated by a surrogate model $\widehat{\boldsymbol{G}}$. In this work, we construct such a surrogate model with a graph neural network, which leverages the geometric relationship between the data points. As such a relationship is already prescribed by the mesh, it is preferable to convert the mesh into a graph for subsequent use with the graph neural network.

As discussed in the introduction, one could transform the mesh into a graph consisting of nodes connected by edges (cf. Fig. \ref{fig:mesh2negraph}), and a generalized graph message-passing network operating on such a graph would mimic the finite volume method. Instead of following this approach, we take inspiration from the data connectivity in the finite element method, and transform the mesh into a hypergraph via a different approach in this work, discussed in detail in the subsequent Sec. \ref{sec:mesh2ne+graph}.

\subsection{From mesh to node-element hypergraph}
\label{sec:mesh2ne+graph}
Consider a bounded spatial domain that is meshed (cf. Fig. \ref{fig:mesh2hypergraph}a). In the finite element method, one treats each cell as an element connecting all the vertices of the cell. Mimicking this behavior, we convert each cell within the mesh into an undirected hyperedge (an ''\emph{element}'') that connects all the vertices of the cell, and converts each of these vertices to a \emph{node}, leading to an undirected hypergraph as shown in Fig. \ref{fig:mesh2hypergraph}b. We further explicitly define the connection between each element and each of the nodes it connects as an undirected \emph{element-node edge}, and this gives us a hypergraph $\mathcal{G}=(\mathcal{V},\mathcal{E}_\square,\mathcal{E}_{v})$, in which $\mathcal{V}$ is the set of all nodes, $\mathcal{E}_\square$ is the set of all elements, and $\mathcal{E}_{v}$ is the set of all element-node edges, as illustrated in Fig. \ref{fig:mesh2hypergraph}c. We coin the name \emph{node-element hypergraph} for such a hypergraph $\mathcal{G}$.

With the node-element hypergraph converted from the mesh, one could proceed to transform the mesh information and system states and attach them as features. These transformations are usually specific to the problem in concern, and we thus delay the discussion of these details to Sec. \ref{sec:model}. In the next subsection, we assume that the system states and mesh information is transformed into some node features, element features, and element-node edge features, and define a message-passing network in the general form.

\begin{rem}
It should be mentioned that it is possible to adopt a directed hypergraph, i.e., assigning a sequence to the nodes connected by the elements, rather than an undirected version. Ma et al. \cite{Ma2022fast}, for example, used a directed hypergraph for the simulation of particulate suspensions. However, to achieve permutation invariance by covering all possible permutations, each undirected hyperedge connecting $k$ nodes has to be converted to $k!$ directed hyperedges. This means that the computational cost can be very high when each hyperedge is connecting more than three nodes (e.g., hypergraphs converted from quadrilateral or hexagonal meshes in 2D, or hypergraphs converted from meshes in 3D). We therefore adopt an undirected hypergraph in this work.
\end{rem}

\subsection{Message-passing layer on node-element hypergraph}
\label{sec:elemmp}
We now proceed to define a graph message-passing network that works with the node-element hypergraph. Taking further inspiration from the finite element method, we design hypergraph message-passing layers that mimic the local stiffness matrix assembly process. In the remaining part of this subsection, we would start from the calculation of the element stiffness matrix for a simple Poisson problem (Eq. \ref{eq:elemstiffsum}), and proceed step-by-step towards the formulation of the hypergraph message-passing layers (Eq. \ref{eq:ne+mp} and Eq. \ref{eq:ne+mpbb}). For the sake of consistency and simplicity in symbols, we write the equations in this subsection and the remaining parts of the paper assuming a 2D quadrilateral mesh and the corresponding converted node-element hypergraph (cf. Fig. \ref{fig:mesh2hypergraph}).

\subsubsection{Finite element form for Poisson equation}
To begin, consider a simple Poisson equation defined on a 2D bounded domain $\Omega$,
\begin{equation}
	-\nabla^2u=F,
\end{equation}
for some scalar $u$ and source term $F$. The weak form of the problem can then be written as
\begin{equation}
	\int_\Omega\nabla w\cdot\nabla u d\Omega=\int_\Omega{wF d\Omega},
\end{equation}
where $w$ being the test function. Discretize the domain with a quadrilateral mesh, the matrix form can be written as
\begin{equation}
	\boldsymbol{K}\boldsymbol{U}=\boldsymbol{F},
\end{equation}
with the stiffness matrix $\boldsymbol{K}$ assembled from the element stiffness matrices $\boldsymbol{K}^e$. We now focus on the element stiffness matrix of an element $\square$ that connects four vertices $i$, $j$, $k$, and $l$. Assume 4-node bi-linear elements and four Gaussian quadrature points, the element stiffness matrix $\boldsymbol{K}_\square^e$ can be summed from the values at individual quadrature points $m=1,2,3,4$,
\begin{equation}
	\label{eq:elemstiffsum}
	\begin{aligned}
	\boldsymbol{K}_\square^e=\sum_{m=1}^4{\boldsymbol{K}_{\square,m}^e}.
	\end{aligned}
\end{equation}
To calculate $\boldsymbol{K}_{\square,m}^e$, we first consider the Jacobian matrix, which can be written as
\begin{equation}
    \boldsymbol{J}_m^e=\left.\frac{\partial(x,y)}{\partial(\xi,\eta)}\right\vert_m=\begin{bmatrix}
    	\frac{\partial x}{\partial \xi}&\frac{\partial x}{\partial \eta}\\
    	\frac{\partial y}{\partial \xi}&\frac{\partial y}{\partial \eta}
    \end{bmatrix},
\end{equation}
where $\xi$ and $\eta$ denote the two axes of the iso-parametric element. With the chain rule, the Jacobian matrix is calculated through
\begin{equation}
	\boldsymbol{J}_m^e=\sum\nolimits_r\left(\begin{bmatrix}
		x_r\\
		y_r
	\end{bmatrix}
\begin{bmatrix}
	\frac{\partial N_{m,r}}{\partial \xi}&\frac{\partial N_{m,r}}{\partial \eta}
\end{bmatrix}\right),
\end{equation}
where $N$ denotes the shape function and subscript $r=i,j,k,l$ denotes the vertex index. Note that the value of the derivatives $\nicefrac{\partial N_{m,r}}{\partial \xi}$ and $\nicefrac{\partial N_{m,r}}{\partial \eta}$ only depend on $m$ and $r$, meaning that if we flatten the Jacobian matrix into a (column) vector, it can be written as
\begin{equation}
	\label{eq:elemstageraw}
	\begin{aligned}
		\operatorname{vec}(\boldsymbol{J}_m^e)&=\sum\nolimits_r(\phi_m^e\left(\begin{bmatrix}
			x_r\\y_r
		\end{bmatrix},r\right))\\
		&=\operatorname{AGG}_r^e(\phi_m^e\left(\begin{bmatrix}
			x_r\\y_r
		\end{bmatrix},r\right)),
	\end{aligned}
\end{equation}
where $\operatorname{vec}$ denotes the flattening operation, the symbol $\phi$ here (and in the rest of this subsection) denotes a function, and $\operatorname{AGG}^e$ denotes an aggregation function which is summation in this case. 

The calculated Jacobian matrix is used to calculated the gradient of the shape function $\nabla N_{m,r}$ at quadrature point $m$
\begin{equation}
	\label{eq:gradn}
	\nabla N_{m,r}=(\boldsymbol{J}_m^e)^{-T}\begin{bmatrix}
		\frac{\partial N_{m,r}}{\partial \xi}\\
		\frac{\partial N_{m,r}}{\partial \eta}
	\end{bmatrix}.
\end{equation}
Since $\boldsymbol{J}_m^e$ is a $2\times2$ matrix, its inverse transpose is a function of its determinant and the value of its individual entries,
\begin{equation}
	\label{eq:jinv}
	\operatorname{vec}\left((\boldsymbol{J}_m^e)^{-T}\right)=\phi_m^{v,1}(\operatorname{det}(\boldsymbol{J}_m^e),\operatorname{vec}(\boldsymbol{J}_m^e)).
\end{equation}
This means that Eq. \ref{eq:gradn} can be rewritten as
\begin{equation}
	\label{eq:gradnfunc}
	\nabla N_{m,r}=\phi_m^{v,2}(\operatorname{det}(\boldsymbol{J}_m^e),\operatorname{vec}(\boldsymbol{J}_m^e),r).
\end{equation}

With the gradient of shape functions $\nabla N_{m,r}$ available, the next step is to calculate the entries of the element stiffness matrix. Note that within the $4\times4$ element stiffness matrix $\boldsymbol{K}_{\square,m}^e$, only a $4\times1$ vector $\boldsymbol{K}_{\square,m,r}^e$ will be assembled to each of the vertices $r=i,j,k,l$, and this vector can be calculated through
\begin{equation}
	\label{eq:elemstiff}
	\boldsymbol{K}_{\square,m,r}^e=\operatorname{det}(\boldsymbol{J}_m^e)\begin{bmatrix}
		(\nabla N_{m,i})^T\\
		(\nabla N_{m,j})^T\\
		(\nabla N_{m,k})^T\\
		(\nabla N_{m,l})^T
	\end{bmatrix}\nabla N_{m,r}.
\end{equation}
Note that the $4\times2$ matrix $\begin{bmatrix}
	\nabla N_{m,i}&\nabla N_{m,j}&\nabla N_{m,k}&\nabla N_{m,l}
\end{bmatrix}^T$ does not depend on $r$ and therefore only a function of the Jacobian matrix $\boldsymbol{J}_m^e$ and its determinant
\begin{equation}
	\label{eq:gradnsfunc}
	\operatorname{vec}\left(
	\begin{bmatrix}
		(\nabla N_{m,i})^T\\
		(\nabla N_{m,j})^T\\
		(\nabla N_{m,k})^T\\
		(\nabla N_{m,l})^T
	\end{bmatrix}\right)=\phi_m^{v,3}(\operatorname{det}(\boldsymbol{J}_m^e),\operatorname{vec}(\boldsymbol{J}_m^e)).
\end{equation}
Combining Eq. \ref{eq:elemstiff} with Eq. \ref{eq:gradnfunc} and \ref{eq:gradnsfunc} gives us
\begin{equation}
	\label{eq:nodestageraw}
	\begin{aligned}
	\boldsymbol{K}_{\square,m,r}^e&=\phi_m^{v,4}\left(\operatorname{det}(\boldsymbol{J}_m^e),\phi_m^{v,3}(\operatorname{det}(\boldsymbol{J}_m^e),\operatorname{vec}(\boldsymbol{J}_m^e)),\phi_m^{v,2}(\operatorname{det}(\boldsymbol{J}_m^e),\operatorname{vec}(\boldsymbol{J}_m^e),r)\right)\\
	&=\phi_m^{v,5}(\operatorname{det}(\boldsymbol{J}_m^e),\operatorname{vec}(\boldsymbol{J}_m^e),r)\\
	&\approx\phi_m^v(S_\square,\operatorname{vec}(\boldsymbol{J}_m^e),r),
    \end{aligned}
\end{equation}
The last step in Eq. \ref{eq:nodestageraw} approximates $\operatorname{det}(\boldsymbol{J}_m^e)$ by the area $S_\square$ of the cell corresponding to the element. Since $\boldsymbol{K}_{\square,r}^e$ is simply a summation of $\boldsymbol{K}_{\square,m,r}^e$ over different quadrature points $m$, we can rewrite Eq. \ref{eq:elemstageraw} and \ref{eq:nodestageraw} as
\begin{subequations}
	\label{eq:kmat}
	\begin{equation}
		\label{eq:kmata}
		\begin{bmatrix}
			\operatorname{vec}(\boldsymbol{J}_1^e)\\
			\operatorname{vec}(\boldsymbol{J}_2^e)\\
			\operatorname{vec}(\boldsymbol{J}_3^e)\\
			\operatorname{vec}(\boldsymbol{J}_4^e)
		\end{bmatrix}=\operatorname{AGG}_r^e(\phi^e\left(\begin{bmatrix}
			x_r\\y_r
		\end{bmatrix},r\right))
	\end{equation}
and
	\begin{equation}
		\label{eq:kmatb}
		\boldsymbol{K}_{\square,r}^e\approx\phi^v(S_\square,\begin{bmatrix}
			\operatorname{vec}(\boldsymbol{J}_1^e)\\
			\operatorname{vec}(\boldsymbol{J}_2^e)\\
			\operatorname{vec}(\boldsymbol{J}_3^e)\\
			\operatorname{vec}(\boldsymbol{J}_4^e)
		\end{bmatrix},r).
	\end{equation}
\end{subequations}

\subsubsection{Conversion to hypergraph message-passing layer}
The next step is to convert Eq. \ref{eq:kmat} into a form that is suitable as a hypergraph neural network layer in general application scenarios. With the node-element hypergraph defined in Sec. \ref{sec:mesh2ne+graph}, we assume that the available system information at each vertex $i$ is converted to node feature $v_i$, system information at each cell $\square$ is converted to the element feature $e_\square$, and the system information at each cell $\square$ that is specific to a certain vertex $r$ connected by it is converted to the element-node edge feature $e_{\square,r}$. Under such assumption, we rewrite Eq. \ref{eq:kmat} in a general form
\begin{subequations}
	\label{eq:ne+mp}
	\begin{equation}
		\label{eq:ne+mpa}
		e_{\square}^\prime=\operatorname{AGG}_r^e\left(\phi^{e}(v_r,e_\square,e_{\square,r})\right),
	\end{equation}
	\begin{equation}
		\label{eq:ne+mpba}
		e_{\square,r}^{\prime\prime}=\left(\phi^{v}(v_r,e_{\square}^\prime,e_{\square,r})\right).
	\end{equation}
\end{subequations}
which is equivalent to Eq. \ref{eq:kmat} by designating
\begin{subequations}
\begin{equation}
	v_r=\begin{bmatrix}
		x_r\\y_r
	\end{bmatrix},\quad\quad e_\square=S_\square,\quad\quad e_{\square,r}=r,
\end{equation}
\begin{equation}
 e_\square^\prime=\begin{bmatrix}
	S_\square\\
	\operatorname{vec}(\boldsymbol{J}_1^e)\\
	\operatorname{vec}(\boldsymbol{J}_2^e)\\
	\operatorname{vec}(\boldsymbol{J}_3^e)\\
	\operatorname{vec}(\boldsymbol{J}_4^e)
\end{bmatrix},
\quad\quad e_{\square,r}^{\prime\prime}=\boldsymbol{K}_{\square,r}^e,
\end{equation}
\end{subequations}
choosing the aggregation function $\operatorname{AGG}^e$ as summation, and specifying that the function $\phi^v$ do not use the additional inputs $v_r$ in Eq. \ref{eq:ne+mpba}.

In the stiffness matrix assembly process in finite element method, the vector $\boldsymbol{K}_{\square,r}^e$ obtained through Eq. \ref{eq:kmatb} for different elements will be assembled to different locations within the stiffness matrix. For the neural network, since the features are usually embedded in high dimensions, we simplify this process to an aggregation rather than assembly
\begin{equation}
	\label{eq:ne+mpbb}
	v_i^\prime=\operatorname{AGG}_\square^v(e_{\square,i}^{\prime\prime}),
\end{equation}
in which $\operatorname{AGG}_\square^v$ denotes an aggregation function over all elements $\square$ that connect node $i$ with other nodes.

\begin{figure}[t]
	\centering
	\includegraphics[]{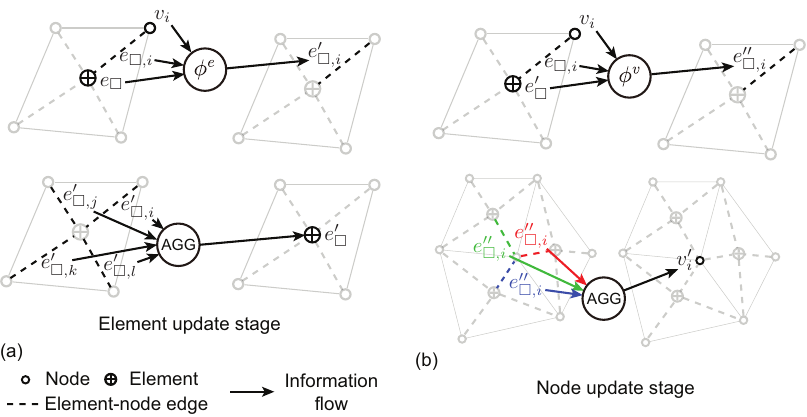}
	\caption{Schematic of the element and node update stages within each hypergraph message-passing layer: (a) element update stage described by Eq.~\ref{eq:ne+mpa}, (b) node update stage described by Eq.~\ref{eq:ne+mpba} and \ref{eq:ne+mpbb}.}
	\label{fig:ne+mp}
\end{figure}

We describe Eq. \ref{eq:ne+mpa} as the \emph{element update stage}, Eq. \ref{eq:ne+mpba} and \ref{eq:ne+mpbb} as the \emph{node update stage}, and write the two stages together as a single message-passing layer
\begin{equation}
	\label{eq:ne+mps}
	v_i^\prime,e_\square^\prime=G_j(v_i,e_{\square,i},e_\square),
\end{equation}
in which the subscript $j$ is used to distinguish between different message-passing layers. An illustration of the two hypergraph message-passing stages described in Eq.~\ref{eq:ne+mp} and Eq.~\ref{eq:ne+mpbb} is plotted in Fig.~\ref{fig:ne+mp}. 

\subsection{Multi-layer perceptrons}
\label{sec:mlp}
The behavior of the element and node update stages described in Eq. \ref{eq:ne+mp} and \ref{eq:ne+mpbb} are largely determined by the choice of functions $\phi^e$ and $\phi^v$. For the purpose of the neural network, we wish to retain the flexibility to learn any dynamics rather than just the local stiffness matrix calculation process, and it is therefore preferable that these functions are universal approximators that can be trained to approximate any function. Under this consideration, we follow the practice shared by most of the works discussed in the introduction, and choose each of these functions $\phi$ as a separate multi-layer perceptron, which can be written as
\begin{equation}
	\phi(\boldsymbol{z}) = f_{n_l}\circ f_{n_l-1}\circ f_{n_l-2}\circ\cdots\circ f_2\circ f_1(\boldsymbol{z}),
\end{equation}
in which the symbol $\circ$ denotes the function composition, and $\boldsymbol{z}$ is the input vector to the multi-layer perceptron. Each layer $f_i$ within the multi-layer perceptron is defined as
\begin{equation}
	f_i(\boldsymbol{z})=\begin{cases}
		\sigma_i(\boldsymbol{W}_i\boldsymbol{z}+\boldsymbol{b}_i) & i=1,2,\ldots,n_l-1\\
		\boldsymbol{W}_i\boldsymbol{z}+\boldsymbol{b}_i & i=n_l\\
	\end{cases}
\end{equation}
where $\boldsymbol{W}_i\boldsymbol{z}+\boldsymbol{b}_i$ is a linear transformation of the input vector $\boldsymbol{z}$, and $\sigma_i$ is a non-linear function called the activation function. The values of all entries within the matrix $\boldsymbol{W}_i$ and the vector $\boldsymbol{b}_i$ are trainable.

\subsection{Network architecture}
\label{sec:network}
With the message-passing layer defined on node-element hypergraph , we are able to construct the surrogate graph neural network model $\widehat{\boldsymbol{G}}$ in Eq. \ref{eq:rom}. Following the encode-process-decode architecture \cite{pfaff2020learning}, the network includes encoders, a series of message-passing layers, and decoders, stacked together in a feedforward fashion. Separate encoders and decoders are constructed for the node features, element-node edge features, and element features respectively using multi-layer perceptrons. The model is summarized in Algorithm \ref{alg:model}.
\begin{algorithm}
	\caption{Finite element-inspired hypergraph neural network ($\phi$-GNN)}
	\label{alg:model}
	\begin{algorithmic}
		\STATE \textbf{Input}: Hypergraph $\mathcal{G}=(\mathcal{V},\mathcal{E}_\square,\mathcal{E}_{v})$, with node features $v_i$, element-node edge features $e_{\square,i}$ and element features $e_{\square}$
		\STATE \textbf{Network}: Node encoder $g^v$, element-node encoder $g^{ev}$, element encoder $g^e$, network layers $G_j$,\\
		node decoder $h^v$, element-node decoder $h^{ev}$, element decoder $h^e$, network depth $n_d$\\
		\STATE \textbf{Encoding}: $v_i\gets g^v(v_i)$, $e_{\square,i}\gets g^{ev}(e_{\square,i})$, $e_\square\gets g^e(e_\square)$
		\STATE $j=0$
		\WHILE{$j<n_d$}
		\STATE $v_i,e_\square\gets G_j(v_i,e_{\square,i},e_\square)$
		\STATE $j\gets j+1$
		\ENDWHILE
		\STATE \textbf{Decoding}: $v_i\gets h^v(v_i)$, $e_{\square,i}\gets h^{ev}(v_i,e_{\square,i},e_\square)$, $e_\square\gets h^e(e_\square)$\\
		\STATE \textbf{Output}: Node features $v_i$, element-node edge features $e_{\square,i}$, element features $e_{\square}$
	\end{algorithmic}
\end{algorithm}

\section{Implementation for fluid dynamics simulations}
\label{sec:model}
In this section, we present a framework that utilizes the $\phi$-GNN defined in Sec. \ref{sec:method} to the surrogate modeling of fluid systems. We would start with the transformation of the mesh and system states into the appropriate features in Sec. \ref{sec:feats}, and then describe the pre-processing (Sec. \ref{sec:preprocess}) and post-processing (Sec. \ref{sec:postprocess}) procedures that are used alongside the network to form the complete framework.

\subsection{Feature transformation, translation invariance and rotation equivariance}
\label{sec:feats}
We consider a fluid system defined on a bounded domain on $\mathbb{R}^2$ with a given mesh, with flow velocity $\boldsymbol{u}=(u_x,u_y)$ and pressure $p$ available at each vertex within the mesh, along with the coordinate of the vertex itself $(x,y)$. We assume that the Reynolds number $\operatorname{Re}$ is known. The mesh is assumed to be body-fitting, i.e., each system boundary is defined as a set of mesh cell boundaries, with boundary conditions ${\Gamma}$ known. The input features to the neural network are transformed from these information. 

Similar to the characteristics of traditional computational fluid dynamics solvers, we expect the network to satisfy a series of invariance and equivariance restrictions. In particular, the predicted pressure values on the vertices are expected to be invariant when the input system translates and/or rotates on the 2D plane. The predicted flow velocity vectors, on the other hand, are not expected to change when the system translates, but are expected to rotate together with the system. These properties are not necessarily satisfied by a graph neural network by default, and therefore the available information $(x,y)$, $(u_x,u_y)$ and $p$ on the vertices need to be transformed to suitable forms before being attached to features.

\begin{figure}[t]
	\centering
	\includegraphics[]{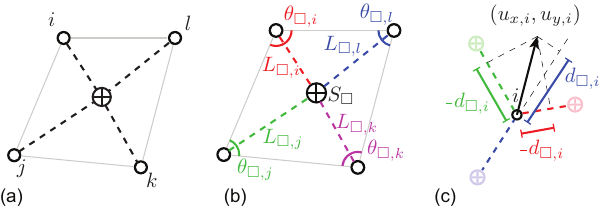}
	\caption{Geometry and flow feature transformations: (a) an element $\square$ connecting nodes $i$, $j$, $k$, $l$, (b) the geometry features used, and (c) projection of flow velocity $(u_{x,i},u_{y,i})$ onto the direction of the local coordinates of the element-node edges.}
	\label{fig:feat}
\end{figure}

\paragraph*{Geometry features}
We first transform the geometry information, namely the coordinates of the vertices, to graph features that are invariant to translation and rotation of the system. This is achieved through converting the \emph{global} coordinate values to \emph{local} geometry features within each element. Given an element (Fig. \ref{fig:feat}a) connecting nodes $i$, $j$, $k$, and $l$, we also give a coordinate to the element $\square$. For simplicity, we designate that this coordinate is at the center of the cell
\begin{equation}
	(x_\square,y_\square)=(\overline{x_r},\overline{y_r}),
\end{equation}
with the overline $\overline{(\cdot)}$ denoting the mean over $r=i,j,k,l$. The local coordinates of the nodes relative to the element can then be written as
\begin{equation}
	(x_{\square,r},y_{\square,r})=(x_r-\overline{x_r},y_r-\overline{y_r}).
\end{equation}
The shape and size of the element can then be fully constrained (in fact over-constrained) by the length of the local coordinate vectors $L_{\square,r}=\sqrt{x_{\square,r}^2+y_{\square,r}^2}$, the angles at the four corners of the element $\theta_{\square,r}$, and the area of the element $S_\square$, as marked out in Fig. \ref{fig:feat}b. These features satisfy the desired translation and rotation invariance properties, and therefore suitable to be attached to features.

\paragraph*{Fluid flow features}
The fluid flow information available are the flow velocity $(u_x,u_y)$ and pressure $p$ at each node, the Reynolds number $\operatorname{Re}$, as well as the set of boundary conditions ${\Gamma}$. With the geometry features being translation and rotation invariant, no specific treatment is needed for the scalar value of pressure. The flow velocity, on the other hand, requires specific treatment. Similar to the approach adopted by Lino et al. \cite{Lino2022} for graph neural network defined on normal graphs, we project the flow velocity vector of each vertex (node) onto the direction of the unit local coordinate vectors $\boldsymbol{x}_{\square,i}$ of the node, 
\begin{equation}
	\label{eq:proj}
	d_{\square,i} = \begin{bmatrix}
		u_{x,i}&u_{y,i}
	\end{bmatrix} \boldsymbol{x}_{\square,i} = \frac{1}{L_{\square,i}}\begin{bmatrix}
		u_{x,i}&u_{y,i}
	\end{bmatrix}\begin{bmatrix}
		x_{\square,i}\\
		y_{\square,i}
	\end{bmatrix},
\end{equation}
as illustrated in Fig. \ref{fig:feat}c. The weights of these projections $d_{\square,i}$ are suitable to be used as features. Following the practice of Pfaff et al. \cite{pfaff2020learning}, we transform the boundary conditions into an one-hot vector $\boldsymbol{\gamma}_i$ for each node $i$.

\subsection{Preprocessing, graph feature attachment}
\label{sec:preprocess}
With the system information transformed to features of suitable form, we proceed to attach them to the graph. Since the system boundaries are body-fitting, the one-hot boundary condition vector is attached to the node feature $v_i$ at each node $i$:
\begin{equation}
	v_i=\boldsymbol{\gamma}_i.
\end{equation}

It is intuitive to attach the element area to the element feature $e_\square$. As the cell area at different regions within the mesh can vary by orders of magnitude, and that the cell areas are strictly greater than zero, we take the natural logarithm of the actual area values,
\begin{equation}
	e_\square=\begin{bmatrix}
		\ln S_\square\\
		-\ln S_\square
	\end{bmatrix},
\end{equation}
in which the $-\ln S_\square$ feature does not provide additional information, and is used to prevent the existence of feature vector of length 1.

The remaining system information are attached to the element-node edge feature. The pressure values $p_i$, originally defined on each node $i$, are gathered to all the element-node edges connected to it,
\begin{equation}
	p_{i\square,i}=p_i,
\end{equation}
in which the subscript $(\cdot)_{i\square,i}$ denotes all the element-node edges that connect node $i$ with an element. Similar to the element area, we also take the natural logarithm values for the length of local coordinate vectors $L_{\square,p}$. Since the mesh enforces that $0<\theta_{\square,i}<\pi$, we can take the cosine of the angles $\theta_{\square,i}$ for the ease of computation and normalization without the loss of information. The element-node edge feature vector, after these treatments, can be written as 
\begin{equation}
	\label{eq:nefeatcyl}
	e_{\square,i}=\begin{bmatrix}
		d_{\square,i}&
		p_{\square,i}&
		\ln L_{\square,i}&
		\cos \theta_{\square,i}
	\end{bmatrix}^T.
\end{equation}
When a variety of dynamics (i.e., a range of Reynolds numbers) are involved, to make it possible for the network to differentiate between similar dynamics, we explicitly supply Reynolds number $\operatorname{Re}$ as a feature, which gives the element-node edge feature vector
\begin{equation}
	\label{eq:nefeataf}
	e_{\square,i}=\begin{bmatrix}
		d_{\square,i}&
		p_{\square,i}&
		\operatorname{Re}&
		\ln L_{\square,i}&
		\cos \theta_{\square,i}
	\end{bmatrix}^T.
\end{equation}
A discussion on the necessity of this treatment is provided in \ref{sec:withre}.

\subsection{Time stepping, postprocessing}
\label{sec:postprocess}
In the temporal roll-out predictions, the network iteratively takes the features as inputs, and updates some of these features with its outputs. Assuming that the boundary conditions and geometry information do not need to be predicted, the features that need to be updated at every time step during the roll-out prediction are the first two entries $d_{\square,i}$ and $p_{\square,i}$ of the element-node edge feature vectors. With the system discretized temporally with forward Euler time-stepping scheme (Eq. \ref{eq:rom}), we have the temporal update from time step $t_n$ to time step $t_{n+1}$:
\begin{equation}
	\label{eq:timestep}
	\begin{aligned}
		e_{\square,i}^{n+1}&=e_{\square,i}^n+\begin{bmatrix}
			\delta d_{\square,i}^n&
			\delta p_{\square,i}^n&
			0&
			0&
			0
		\end{bmatrix}^T\\
		&\approx e_{\square,i}^n+\begin{bmatrix}
			\widehat{\boldsymbol{G}}(v_i,e_{\square,i}^n,e_\square)\\
				\boldsymbol{0}_{3\times1}
		\end{bmatrix},
	\end{aligned}
\end{equation}
with $\widehat{\boldsymbol{G}}$ being the $\phi$-GNN defined in Sec. \ref{sec:network}. The network outputs are vectors of length 2 on element-node edges, and the decoders for the node and element features $h^v$ and $h^e$ are not used.

It should be noted that the temporal update are on the element-node edges which do not represent a physical location. To retrieve the velocity $(u,v)$ and pressure $p$ fields defined on vertices of the mesh, we aggregate the element-node edge features back to the nodes as a post-processing step. For the pressure $p$, the value on each node can be simply retrieved by calculating the mean value across all the element-node edges connecting to the node
\begin{equation}
	p_i^n=\operatorname{MEAN}_\square(p_{\square,i}^n),
\end{equation}
in which $\operatorname{MEAN}_\square$ is the mean aggregation function. The velocity at each node is retrieved by reverting the process of Eq. \ref{eq:proj}. Similar to that in Lino et al. \cite{Lino2022}, this can be achieved by solving the (usually over-constrained) problem via Moore-Penrose pseudo-inverse:
\begin{equation}
	\label{eq:invgeoproj}
	\boldsymbol{u}_i^n=\begin{bmatrix}
		u_{x,i}^n\\
		u_{y,i}^n\\
	\end{bmatrix}=\left(\boldsymbol{d}_{i\square,i}^n\odot((\boldsymbol{X}_{i\square,i}\boldsymbol{X}_{i\square,i}^T)^{-1}\boldsymbol{X}_{i\square,i})\right)\boldsymbol{I}_{n_{i\square,i}\times1},
\end{equation}
in which the symbol $\odot$ denotes the Hadamard product, $n_{i\square,i}$ denotes the number of element-node edges connecting to node $i$, $\boldsymbol{d}_{i\square,i}$ is a $2\times n_{i\square,i}$ matrix containing the projection weights $d_{i\square,i}$ for all the element-node edges connecting to node $i$, and $\boldsymbol{X}_{i\square,i}$
is also a $2\times n_{i\square,i}$ matrix containing the unit local coordinate vectors $\boldsymbol{x}_{\square,i}$ for all the element-node edges connecting to node $i$.

\begin{rem}
	In the case when a node $i$ only connects to one element (e.g., at the edge in domain boundary), Eq. \ref{eq:invgeoproj} cannot be solved due to the problem being under-constrained, and we simply set $\boldsymbol{d}_{i\square,i}^n\equiv0$ and $\widehat{\boldsymbol{u}}_i^n\equiv0$.
\end{rem}

\section{Problem Set-up}
\label{sec:exp}
In this section, we apply the framework developed in Sec. \ref{sec:model} to the modeling and prediction of two fluid systems. We will first describe the data sets used, and then discuss the details on the setup and training of the models. The results will be presented in Sec. \ref{sec:results}. All neural network trainings are performed with random seeds fixed at 1 unless specifically mentioned.

\subsection{Data sets}
We choose two common benchmark fluid flow systems for the experiments: The flow around a circular cylinder and the flow around an NACA0012 airfoil. Both flow data sets are 2-D incompressible fluid flow simulated at laminar flow conditions. The flow around the cylinder data set is obtained through simulation at Reynolds number $Re=200$, and used to test the capability of the neural network in learning a certain flow dynamic. The flow around the airfoil is simulated at multiple Reynolds numbers (with the flow at each Reynolds number forming a separate "trajectory") within the range $Re\in[1000,4000]$, and the resulting data set is used to test the capability of the network to interpolate within a range of dynamics and extrapolate out of the range.

\begin{figure}[t]
	\centering
	\includegraphics[]{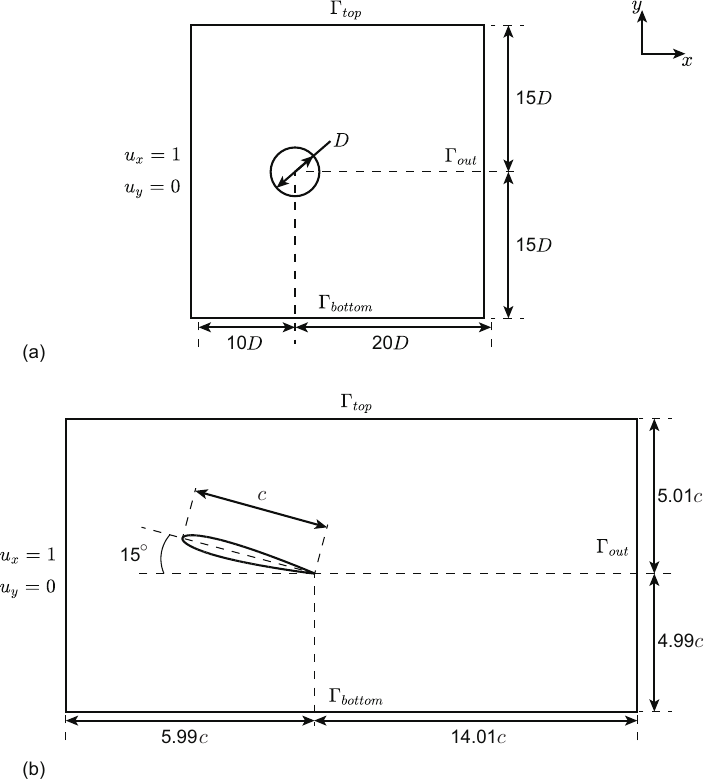}
	\caption{Schematic of the computational domain for the two data sets: (a) flow around a circular cylinder (b) flow around a NACA0012 airfoil at 15 degrees angle of attack.}
	\label{fig:domain}
\end{figure}

\begin{figure}[h]
	\centering
	\includegraphics[]{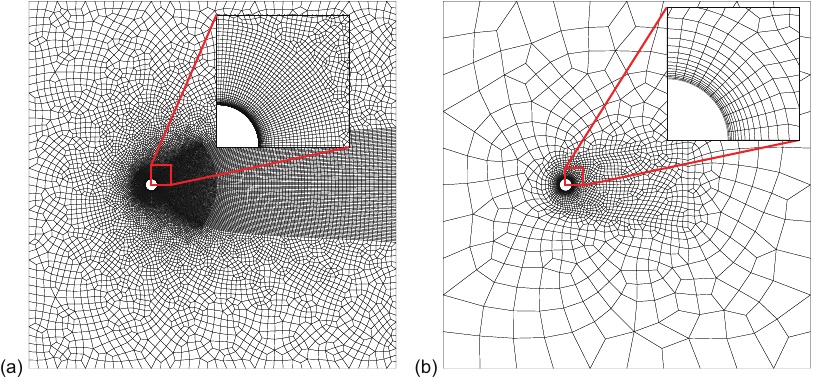}
	\caption{Schematic of the mesh used for the flow around cylinder case: (a) computational mesh for the CFD solver, (b) coarser mesh to be converted to a hypergraph for the graph neural network.}
	\label{fig:cylmesh}
\end{figure}

\begin{figure}[t]
	\centering
	\includegraphics[]{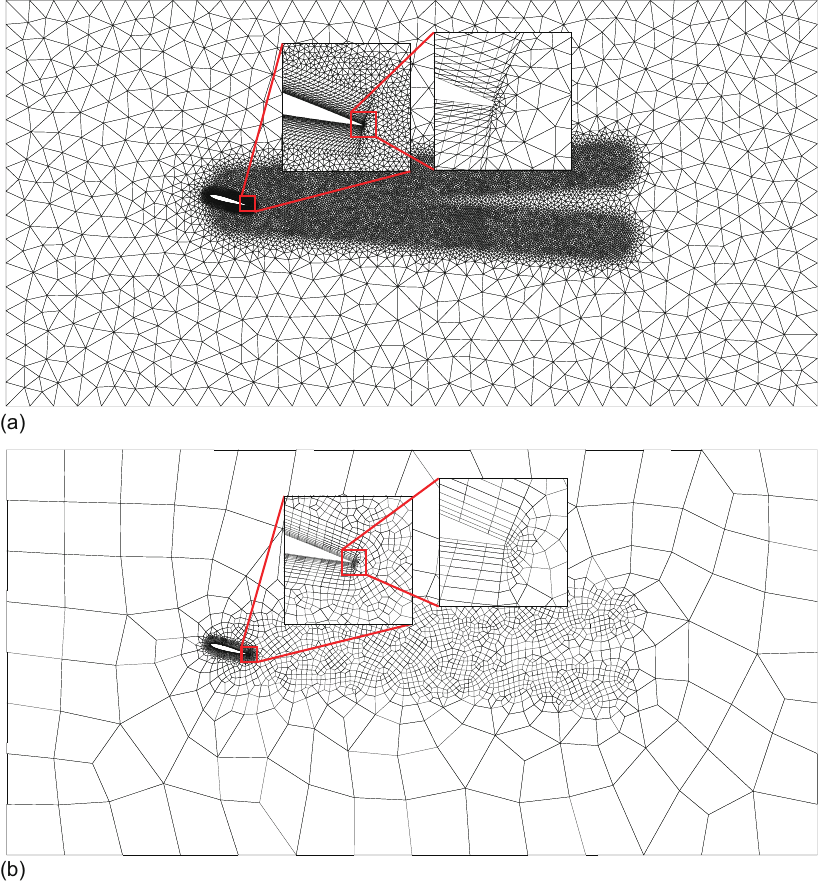}
	\caption{Schematic of the mesh used for the flow around airfoil case: (a) computational mesh for the CFD solver, (b) coarser mesh to be converted to a hypergraph for the graph neural network.}
	\label{fig:airfoilmesh}
\end{figure}

Both flow data sets are generated via a Petrov-Galerkin finite element solver with a semi-discrete time-stepping scheme \cite{Jaiman2016} written in Matlab, with the computational meshes created using Gmsh \cite{geuzaine2009gmsh}. The domain for the two sets of simulations is plotted in Fig. \ref{fig:domain}. The inlet is a fixed uniform flow $u_x=1, u_y=0$, the outlet $\Gamma_{out}$ is set to be traction free, while top and bottom of the boundary conditions $\Gamma_{top}$ and $\Gamma_{bottom}$ are set to be slip-wall. No slip boundary condition is applied at the surface of the cylinder and airfoil. The computational meshes are shown in Fig. \ref{fig:cylmesh}a and Fig. \ref{fig:airfoilmesh}a. For the flow around cylinder, a total of 6499 continuous time steps are sampled with non-dimensionalized time step $\Delta t^*=0.04$ at Reynolds number $Re=200$. For the flow around an airfoil, simulations are performed at 61 different Reynolds numbers within the range $Re\in[1000,4000]$. For each of the considered Reynolds number, 4500 continuous time steps are sampled with non-dimensionalized time step $\Delta t^*=0.0167$. 

The simulated flow data are interpolated onto a coarser mesh (depicted in Fig. \ref{fig:cylmesh}b and \ref{fig:airfoilmesh}b) via a Clough-Tocher interpolator available in SciPy package \cite{Virtanen2020} before subsequent conversion to features. It should be noted that the mesh for the flow around fixed cylinder is the same as the one used for the fluid-structure interaction between fluid flow and an elastically mounted cylinder in reference \cite{Gao2022quasi}. The statistics about the number of nodes, elements and element-node edges for the two data sets are reported in Table \ref{tab:nodeedgecount}. The interpolated data sets are then split into train, cross-validation and test data sets, as listed in Table \ref{tab:traintestsets}. It should be noted that the training and cross-validation share the same data set for the cylinder case. This is possible because the network is trained with single-step supervision, but evaluated by generating roll-out predictions for thousands of time steps starting from the first time step within the training/cross-validation data set.

\begin{table}[h]
	\caption{Statistics of the converted graph/hypergraph for the data sets used}
	\centering
	\begin{tabular}{cccc}
		\toprule
		Data set & Nodes & Elements & Element-node edges \\
		\midrule
		Cylinder	& 2204 & 2160 & 8640 \\
		Airfoil	& 3653 & 3590 & 14360 \\
		\bottomrule
	\end{tabular}
	\label{tab:nodeedgecount}
\end{table}

\begin{table}[h]
	\caption{Train-test split of the data sets. The $a_1:b:a_2$ notation denotes a series of values starting from $a_1$ and ends at $a_2$, with interval $b$ and both ends included.}
	\centering
	\begin{tabular}{ccccc}
		\toprule
		Data set & \begin{tabular}[c]{@{}c@{}}Training \& cross-validation \\Reynolds number\end{tabular} & \begin{tabular}[c]{@{}c@{}}Testing\\Reynolds number\end{tabular} & \begin{tabular}[c]{@{}c@{}}Time steps per\\training trajectory\end{tabular} & \begin{tabular}[c]{@{}c@{}}Time steps per\\testing trajectory\end{tabular} \\
		\midrule
		Cylinder	& 200 & 200 & 2048 & 4000 \\
		Airfoil	& \begin{tabular}[c]{@{}c@{}}2000:100:3000\\2033:100:2933\\2067:100:2967\end{tabular} & \begin{tabular}[c]{@{}c@{}}1000:100:1900\\2050:100:2950\\3100:100:4000\end{tabular} & 256 & 4000 \\
		\bottomrule
	\end{tabular}
	\label{tab:traintestsets}
\end{table}

\subsection{Network setup, implementation, training}
\label{sec:setup}
We implement the models with PyTorch \cite{paszke2019}. All multi-layer perceptrons in the network (encoders, decoders, processors) have two hidden layers with hidden and output layer width 128, except for the outputs of the element-node decoder which have width 2. A total of $n_d=15$ message-passing layers are used. It should be noted that the hyperparameters of network width and depth are not tuned but fixed following the choice of Pfaff et al. \cite{pfaff2020learning} for their MeshGraphNet. Residual links \cite{he2016} are added in all message-passing layers. 

Unlike most of the existing works on graph neural network-based continuum mechanics simulations discussed in the introduction, we adopt a sinusoidal activation function \cite{Sitzmann2020} in this work along with the associated initialization scheme. To avoid conflict with this design choice, we choose to use the mean aggregation function for both the element and node update stages, and avoid the use of any normalization layers. In addition, we apply min-max normalization with minimum $-1$ and maximum $1$ to each entry for each of the input and output features in the training data sets, with the exception of $\cos\theta_{\square,i}$ in the element-node edge features since its value is already limited to be within the range of $(-1,1)$. The test data sets are normalized with the statistics calculated from the training data sets.

Models reported in Sec. \ref{sec:resultscyl} and \ref{sec:resultsaf} are trained with Adam \cite{Kingma2014} optimizer with PyTorch default setup for a total of 200 epochs with batch size 4, with an adaptive smooth $L_1$ training loss that will be discussed in Sec. \ref{sec:loss}. The network is trained using single-step supervision, i.e., for the input features at time step $t_n$, the training loss is calculated between the predicted increment of element-node edge features from time step $t_n$ to time step $t_{n+1}$ (cf. Eq. \ref{eq:timestep}) and the ground truth values. Artificial noise is added to the ground truth input and output values during training, with the scheme discussed in detail later in Sec. \ref{sec:noise}. A maximum learning rate of $10^{-4}$ and a minimum learning rate $10^{-6}$ are selected, with a warm-up stage of 10 epochs at the start. Details of the learning rate scheme will be presented in Sec. \ref{sec:lr}. All the reported training and testing runs are completed on a single Nvidia RTX 3090 GPU with CPU being AMD Ryzen 9 5900 @ 3 GHz $\times$ 12 cores. 

\begin{rem}
	It should be noted that the message-passing layers are implemented via a gather-scatter scheme similar to that of Pytorch Geometric \cite{Fey2019}, which is non-deterministic due to the use of GPU atomic operations. This means that the results from multiple trials of training and testing runs can still be different from each other even when the random seeds are fixed as the same value. Fully-deterministic implementation is possible via the direct use of sparse matrix multiplications, but seems to be significantly slower than the non-deterministic implementation. We therefore adopt the non-deterministic version in the present work.
\end{rem}

\subsection{Adaptive smooth $L_1$ training loss}
\label{sec:loss}
Most of the existing works discussed in Sec. \ref{sec:intro} adopt mean squared error (MSE) as the training loss function. For graph or hypergraph message-passing networks with sinusoidal activation functions, however, we observe that the training process can be unstable when MSE loss is used. We therefore seek an alternative training loss in this work. Another typical loss function, the $L_1$ loss, might not be preferable in this case since it is not smooth at zero, and we thus adopt a smooth $L_1$ loss \cite{girshick2015fast} function, which states that for ground truth $\psi$ and its neural network prediction $\hat{\psi}$, the loss 
\begin{equation}
	\begin{aligned}
		L_i(\psi_i,\hat{\psi}_i) = 
		\begin{cases}
			(\psi_i-\hat{\psi}_i)^2/2\beta,&\text{if }\lvert\psi_i-\hat{\psi}_i\rvert<\beta\\
			\lvert\psi_i-\hat{\psi}_i\rvert-\beta/2,  &\text{if }\lvert\psi_i-\hat{\psi}_i\rvert\geq\beta\\
		\end{cases}
	\end{aligned}
\end{equation}
in which $\beta$ is a non-negative parameter that controls the transition point between the $L_2$ loss region and the $L_1$ loss region. The subscript $(\cdot)_i$ here denotes entry-by-entry calculation. A fixed $\beta$ value is not preferable, since the loss function is not very different from the $L_1$ loss when $\beta$ is too small and will converge to a scaled $L_2$ loss during training when $\beta$ is too large. For the present case, the instability in training usually occurs when the training MSE error is relatively low, so an adaptive scheme for $\beta$ is preferred. Assuming that the distribution of error during the training approximately follows a symmetric distribution centered at zero, the target is to make sure that the two tails of the distribution fall into the $L_1$ loss region so that they do not lead to instability. More complex on-the-fly $\beta$ adaptation algorithms like references \cite{fu2019retinamask,zhang2020dynamic,sutanto2021novel} exist, but we choose to adapt the value $\beta$ in this work using a simpler yet robust approach by setting $\beta^2$ as the variance of the model prediction error
\begin{equation}
	\beta^2=\operatorname{Var}(\psi-\hat{\psi})\approx\operatorname{MSE}(\psi_{train},\hat{\psi}_{train})
\end{equation}
based on the idea that at least part of the two tails of any zero-centered symmetric distribution would reside more than one standard deviation away from zero. This variance of prediction error can be approximated by computing the mean-squared error between the predicted and ground truth value of the whole training set, which can be further approximated on-the-fly by an exponential moving average
\begin{equation}
	\beta^2\leftarrow (1-\frac{1}{N_{b}})\beta^2+\frac{1}{N_{b}}\operatorname{MSE}(\psi_{batch},\hat{\psi}_{batch}),
\end{equation}
in which $N_b$ is the total number of training iterations within each epoch. We further stabilize this process by preventing $\beta$ from increasing, i.e.,
\begin{equation}
	\beta^2\leftarrow (1-\frac{1}{N_{b}})\beta^2+\frac{1}{N_{b}}\min\{\beta^2,\operatorname{MSE}(\psi_{batch},\hat{\psi}_{batch})\}
\end{equation}
for each training step. The initial value of $\beta$ can be determined by calculating the mean-squared error of the initialized network over a small randomly-sampled subset of the whole training set. Such a subset contains $400$ samples for the reported cases in Sec. \ref{sec:results}.

\subsection{Training noise}
\label{sec:noise}
The prediction of the network is expected to have an error. Without special treatment, this error would accumulate over the prediction time steps and would eventually cause the predictions to blow up. It is therefore preferable that the network automatically corrects its prediction error. In this work, this is achieved by training the network with artificial noise added to the training data. With a the trained network $\widehat{G}$, we consider its prediction output $\widehat{\delta e}_{\square,i}^n$ calculated from the input features at time step $t_n$,
\begin{equation}
	e_{\square,i}^{n+1}\approx e_{\square,i}^n+\begin{bmatrix}
		\boldsymbol{\kappa}\\
		\boldsymbol{0}_{3\times1}
	\end{bmatrix}\odot\begin{bmatrix}
		\widehat{\boldsymbol{G}}(v_i,e_{\square,i}^n,e_\square)\\
		\boldsymbol{0}_{3\times1}
	\end{bmatrix}=e_{\square,i}^n+\begin{bmatrix}
	\boldsymbol{\kappa}\\
	\boldsymbol{0}_{3\times1}
\end{bmatrix}\odot\begin{bmatrix}
	\widehat{\delta e}_{\square,i}^n\\
	\boldsymbol{0}_{3\times1}
\end{bmatrix}.
\end{equation}
It should be noted that this equation is different from Eq. \ref{eq:timestep}. The additional scaling factor (vector) $\boldsymbol{\kappa}$ exists due to the necessary normalization of neural network output targets during the training process. Assuming that the network is large enough and trained thorough enough, such that it is able to capture the evolution of the system dynamics, it is reasonable to assume that the error of this prediction would approximately follow a Gaussian distribution,
\begin{equation}
	\widehat{\delta e}_{\square,i}^n-\delta e_{\square,i}^n=\epsilon_{\square,i}^n\sim \mathcal{N}(0,\operatorname{Var}(\epsilon_{\square,i}^n)).
\end{equation}
We hope that the network will be able to correct this error during its prediction in the subsequent time step, i.e., we expect that
\begin{equation}
	\widehat{\boldsymbol{G}}(v_i,e_{\square,i}^{n+1}+\begin{bmatrix}
		\boldsymbol{\kappa}\odot\epsilon_{\square,i}^n\\
		\boldsymbol{0}_{3\times1}
	\end{bmatrix},e_\square)=(\delta e_{\square,i}^{n+1}-\epsilon_{\square,i}^n)+\epsilon_{\square,i}^{n+1},
\end{equation}
in which $\epsilon_{\square,i}^{n+1}\sim\mathcal{N}(0,\operatorname{Var}(\epsilon_{\square,i}^{n+1}))$ is the network prediction error at time step $t_{n+1}$. Therefore, during training, one can enforce the network to learn such a behavior by adding a Gaussian noise $-\widehat{\epsilon}_{\square,i}^{n}$ with a specified variance to the training output target, and $\boldsymbol{\kappa}\odot\widehat{\epsilon}_{\square,i}^n$ to the corresponding entries of training input. 

\begin{rem}
	It should noted that this scheme is directly inspired by but not equivalent to the one adopted by Pfaff et al. \cite{pfaff2020learning}, as they chose to specify a certain magnitude of Gaussian noise for input rather than for the output. Two schemes are only equivalent when different entries of vector $\kappa$ are equal to each other.
\end{rem}

For systems that only changes slightly over each time step, i.e., both entries of $\kappa$ is much lower than 1, the predicted output at neighboring time steps would be similar to each other. It is therefore also reasonable to expect that the prediction error of two neighboring time steps are significantly correlated,
\begin{equation}
	\frac{\epsilon_{\square,i}^n}{\|\epsilon_{\square,i}^n\|}\cdot\frac{\epsilon_{\square,i}^{n+1}}{\|\epsilon_{\square,i}^{n+1}\|}\sim \mathcal{O}(1).
\end{equation}
This is empirically verified during the preliminary tests, we skip the details for brevity. Based on such an observation, we further propose to train the network to over-correct the error at each time step, such that a proportion of error at the subsequent time step is also canceled. This means that for a Gaussian noise $-\widehat{\epsilon}_{\square,i}^{n}$ with a specified variance, we add $-\omega\widehat{\epsilon}_{\square,i}^{n}$ to the training output target, and add $\boldsymbol{\kappa}\odot\widehat{\epsilon}_{\square,i}^n$ to the corresponding entries of input, with $\omega>1$ being a hyper-parameter that controls the level of over-correction. We take inspiration from the typical relaxation factor range used in successive over-relaxation methods, and directly specify $\omega=1.2$ in this work rather than tuning it as a hyper-parameter.

For the flow around cylinder data set, the model reported in Sec. \ref{sec:resultscyl} is trained with the standard deviation of random noise designated to be $6.5\times10^{-2}$. This value is selected after a screen through the range $[10^{-2},10^{-1}]$, and a discussion of the behavior of the network when trained with different amount of training noise is attached in \ref{sec:noisemag}. For the flow around airfoil data sets, to reduce the amount of computational power needed, we start from the noise standard deviation level of $10^{-1}$ and divide this value by half in each trial, eventually using the noise standard deviation level $2.5\times10^{-2}$ for the results reported in Sec. \ref{sec:resultsaf}.

\subsection{Learning rate}
\label{sec:lr}
We choose to merge the popular cosine and exponential learning rate schemes to take the advantage of both of them, such that a scaled cosine learning rate applies near the start of the training and a scaled exponential learning rate applies near the end. In particular, a decaying learning rate curve starting from maximum learning rate $\zeta_{max}$ and ends at minimum learning rate $\zeta_{min}$ is specified by
\begin{subequations}
	\begin{equation}
		\zeta=\begin{cases}
			\zeta_{max}+\delta_{\zeta}\left((1+\cos(\pi\alpha))(\zeta_{max}-\zeta_{min})/2+\zeta_{min}-\zeta_{max}\right), &\text{if }\alpha<\alpha_0\\
			\zeta_{min}+\delta_{\zeta}(\zeta_{max}(\zeta_{min}/\zeta_{max})^\alpha-\zeta_{min}), &\text{if }\alpha>\alpha_0
		\end{cases}
	\end{equation}
	in which the scaling factor
	\begin{equation}
		\delta_\zeta= \frac{\zeta_{max}-\zeta_{min}}{\zeta_{max}-2\zeta_{min}-(1+\cos(\pi\alpha_0))(\zeta_{max}-\zeta_{min})/2+\zeta_{max}(\zeta_{min}/\zeta_{max})^{\alpha_0}}
	\end{equation}
	scales the two parts of the curve such that the curve is continuous over $\alpha\in[0,1]$.
	The point $\alpha_0$ where the gradient of the two parts of the curve matches with each other
	\begin{equation}
		\pi\sin(\pi\alpha_0)\frac{\zeta_{max}-\zeta_{min}}{2}=\zeta_{max}\ln\left(\frac{\zeta_{min}}{\zeta_{max}}\right)(\zeta_{min}/\zeta_{max})^{\alpha_0}
	\end{equation}
is numerically solved starting from a guess $\alpha_0=0.27$.
\end{subequations}
The value $\alpha$ is equal to the proportion of training completed. An increasing learning curve from $\zeta_{min}$ to $\zeta_{max}$, used in the warm up stage at the start of the training process, is acquired by calculating the decaying learning curve and then reverse the sequence. For the models reported in this work, we change the learning rate at the start of every iteration during the warm-up stage, and change the learning rate at the start of every epoch for the rest of the training process.

\subsection{Evaluation metrics}
\label{sec:metric}
The trained models are evaluated on the test data sets. As the main purpose of the neural network surrogate model is to generate stable and accurate roll-out simulations, we evaluate the models by feeding the state of the system at a certain time step, i.e. the first time step within each of the test data sets, to the model and compare the predicted system states over the next several thousands of time steps with the ground truth values from the interpolated CFD data. We evaluate the model based on the following criteria:
\paragraph*{Accuracy \& stability of predicted system states} We quantify the accuracy of the predicted system states by calculating the similarity between the ground truth system state vector $\boldsymbol{q}$ and its neural network prediction $\hat{\boldsymbol{q}^*}$ over the prediction roll-out time steps by calculating the coefficient of determination
\begin{equation}
	R^2=1-\frac{\|\boldsymbol{q}^*-\hat{\boldsymbol{q}^*}\|_2^2}{\|\boldsymbol{q}^*-\overline{\boldsymbol{q}^*}\|_2^2},
\end{equation}
in which $\|\cdot\|_2$ denotes the $L_2$ norm. The system state vector $\boldsymbol{q}$ could be either the velocity components $\boldsymbol{u}_x$ or $\boldsymbol{u}_y$, or the non-dimensionalized pressure field $\boldsymbol{p}^*$. A higher coefficient of determination indicates a more accurate prediction, up to $R^2=1$ which means perfectly accurate predictions. It is also preferable if the network is able to generate stable predictions continuously. While it could be difficult to prove the stability mathematically, it is possible to empirically verify it by using the network to predict over a very long period into the future and compare with the ground truth CFD predictions. In this work, we deem that being able to generate a stable roll-out prediction over 4000 time steps (equivalent to about 30-40 vortex shedding cycles for the test data sets used in this work) is a sufficient empirical verification of stability.

\paragraph*{Accuracy of lift and drag coefficients} We also calculate the lift and drag coefficients directly from the predicted system states. These statistics are compared with the ground truth values calculated from CFD data. In particular, the lift and drag coefficients are calculated by directly integrating the Cauchy stress tensor $\boldsymbol{\sigma}^f=-p^f\boldsymbol{I}+\mu^f(\nabla\boldsymbol{u}+(\nabla\boldsymbol{u})^T)$ at the first layer of mesh away from the surface of the cylinder or airfoil,
\begin{subequations}
	\label{eq:liftdrag}
	\begin{equation}
		C_l=\frac{1}{\frac{1}{2}\rho^f(U_\infty)^2L_c}\int_{\boldsymbol{\Gamma}^{fs}}(\boldsymbol{\sigma}^f\cdot \boldsymbol{n})\cdot \boldsymbol{n}_y d\boldsymbol{\Gamma},
	\end{equation}
	\begin{equation}
		C_d=\frac{1}{\frac{1}{2}\rho^f(U_\infty)^2L_c}\int_{\boldsymbol{\Gamma}^{fs}}(\boldsymbol{\sigma}^f\cdot \boldsymbol{n})\cdot \boldsymbol{n}_x d\boldsymbol{\Gamma},
	\end{equation}
\end{subequations}
where $\rho^f$ is the density of the fluid, $\mu^f$ is the viscosity of the fluid, $U_\infty=1$ is equal to the magnitude of inlet velocity, and $\boldsymbol{\Gamma}^{fs}$ denotes the surface of the airfoil or cylinder. The characteristic length $L_c$ is either equal to the diameter of the cylinder $D$ or the chord length of the airfoil $c$.

It should be reiterated that the lift and drag coefficients reported in this work are not directly given by the neural network, but rather calculated from the predicted velocity $\boldsymbol{u}_x$, $\boldsymbol{u}_y$ and pressure $p$, without any correction or interpolation schemes like the one used in \cite{Gupta2022b}.

\section{Results and discussion}
\label{sec:results}

\subsection{Learning a certain fluid flow dynamic}
\label{sec:resultscyl}
Starting from the first time step in each of the test data sets, we generate roll-out predictions of the states of the fluid system over the next 4000 time steps. In this subsection, we present these results for the flow around cylinder data set, using the criteria discussed in Sec. \ref{sec:metric}. We also verify the designed invariance and equivariance properties of the network.

\begin{figure}
	\centering
	\includegraphics[]{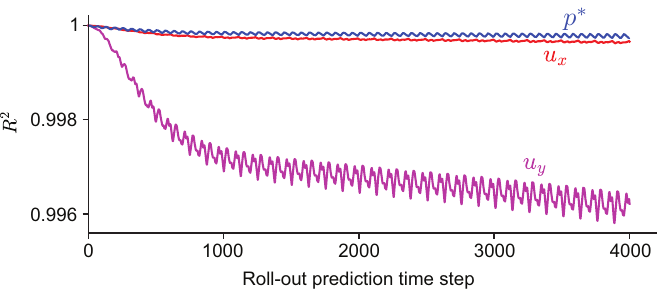}
	\caption{Coefficient of determination for the predicted velocity components $\boldsymbol{u}_x$ and $\boldsymbol{u}_y$, as well as the non-dimensionalized pressure field $\boldsymbol{p}^*$, over 4000 roll-out prediction time steps, starting from the system state at the first time step within the flow around cylinder test data set.}
	\label{fig:r2cyl}
\end{figure}

\begin{figure}
	\centering
	\includegraphics[]{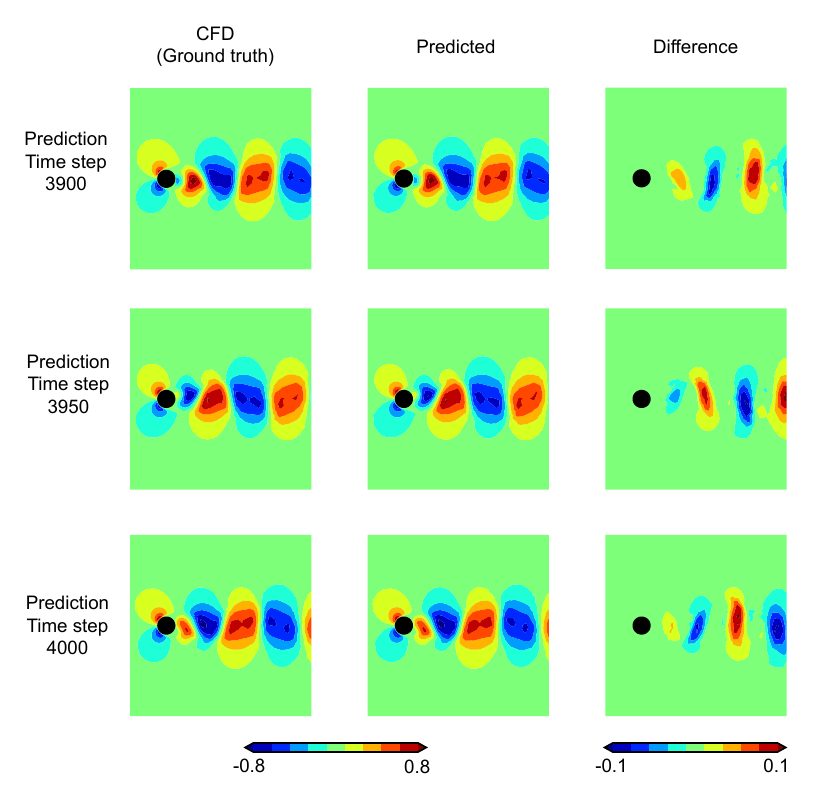}
	\caption{Predicted versus ground truth $y$-axis velocity component $\boldsymbol{u}_y$ at roll-out prediction time steps 3900, 3950 and 4000, zoomed in to the wake of the cylinder.}
	\label{fig:pstarcyl}
\end{figure}

\paragraph*{Field prediction and stability} Figure \ref{fig:r2cyl} shows the coefficient of determination of the predicted systems states $\boldsymbol{u}_x$, $\boldsymbol{u}_y$ and $\boldsymbol{p}^*$ for the flow around cylinder data set over the roll-out prediction over 4000 time steps, and Fig. \ref{fig:pstarcyl} presents the predicted versus ground truth $y$-axis velocity component $\boldsymbol{u}_y$ at time step 3900, 3950 and 4000. An $R^2$ value consistently close to $1$, along with the close match between predicted and ground truth $\boldsymbol{u}_y$, indicate that the predictions are stabilized and accurate.

\begin{figure}
	\centering
	\includegraphics[]{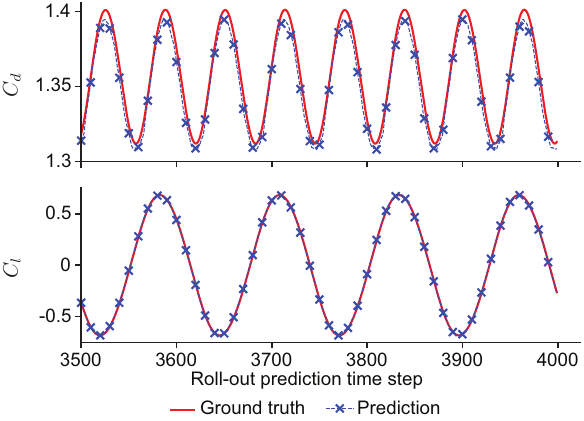}
	\caption{Predicted versus ground truth lift and drag coefficients between prediction time steps 3501 and 4000 for the flow around cylinder case.}
	\label{fig:liftdragcyl}
\end{figure}

\paragraph*{Lift and drag} Using Eq. \ref{eq:liftdrag}, we calculate the lift and drag coefficients from the predicted system states over time, and compare with the ground truth values. The results between roll-out prediction time steps 3501 and 4000 are plotted in Fig. \ref{fig:liftdragcyl}. The lift and drag coefficients calculated from the predicted system states are reasonably accurate, with drag coefficients slightly ($\approx1\%$) lower than the ground truth values.

\begin{figure}
	\centering
	\includegraphics[]{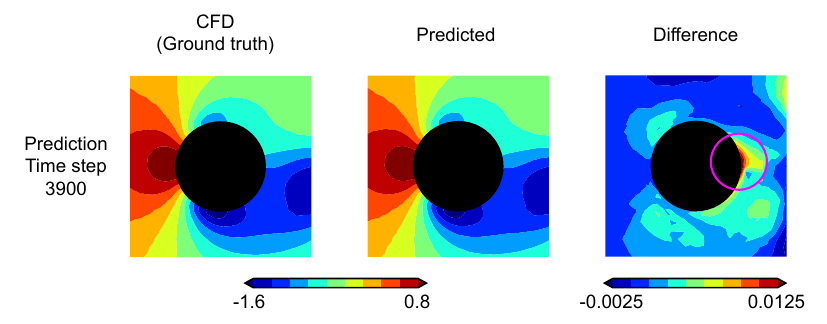}
	\caption{Predicted versus ground truth non-dimensionalized pressure $\boldsymbol{p}^*$ at roll-out prediction time step 3900, zoomed in to the vicinity of the cylinder. The predicted pressure field is slightly higher than the ground truth values in the region marked out by the magenta circle, which is consistent with the fact that the calculated drag forces are slightly lower than the ground truth values in Fig. \ref{fig:liftdragcyl}.}
	\label{fig:pcylzoomin}
\end{figure}

\paragraph*{Predictions near the cylinder surface} We further zoom in to the vicinity of the cylinder, and plot the predicted versus ground truth pressure fields in Fig. \ref{fig:pcylzoomin}. As marked out by a magenta circle, the predicted pressure values around a region near the surface of the cylinder are slightly higher than the ground truth values. Since the lift and drag forces are only integrated using the values at the first layer of mesh away from the cylinder surface, this error in the predicted pressure field would cause the drag coefficient calculated from the predicted system states to be lower than that calculated from the ground truth CFD data, which is consistent with the observations in Fig. \ref{fig:liftdragcyl}.

\begin{figure}
	\centering
	\includegraphics[]{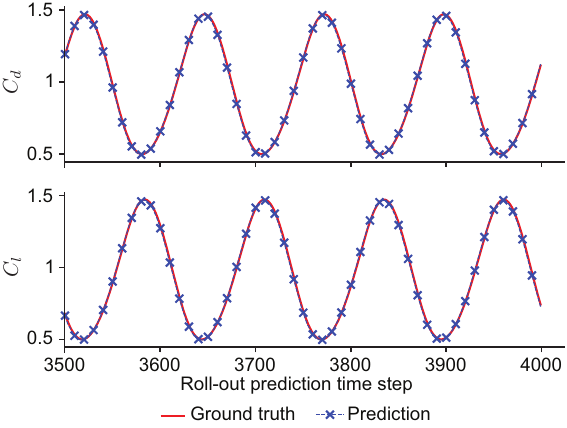}
	\caption{Predicted versus ground truth lift and drag coefficients between prediction time steps 3501 and 4000 for the flow around cylinder case. The system at the initial time step is rotated by $\pi/4$ and translated by a random value before being sent into the neural network.}
	\label{fig:liftdragcylrot}
\end{figure}

\paragraph*{Rotation equivariance and translation invariance} We verify the desired invariance and equivariance properties by rotating and translating the input system before sending it into the network for roll-out predictions. In particular, we rotate the input system at the first time step within the flow around cylinder test data set by $\pi/4$, and then translate the system by a random value. The resulting lift and drag coefficient curves plotted in Fig. \ref{fig:liftdragcylrot} are accurate, showing that these desired properties are satisfied.

\begin{figure}
	\centering
	\includegraphics[]{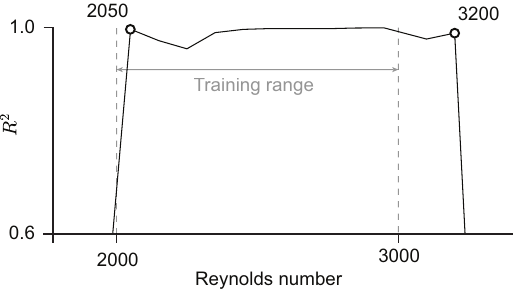}
	\caption{Coefficient of determination of the predicted pressure field $\boldsymbol{p}^*$ at roll-out prediction time step 4000 for different test data sets. The network is able to generate stable predictions for test data sets with Reynolds numbers between 2050 and 3200, and not able to do so for test data sets with Reynolds numbers lower or equal to 1950 as well as test data sets with Reynolds numbers higher or equal to 3300.}
	\label{fig:afstability}
\end{figure}

\subsection{Learning a range of fluid flow dynamics}
\label{sec:resultsaf}
With the network shown to be able to generate stable and accurate predictions when used to learn a single fluid flow dynamics in Sec. \ref{sec:resultscyl}, we proceed to apply the network to learning a range of dynamics using the flow around airfoil data sets, and verify whether it can serves as an effective surrogate model for the time series predictions within this range of Reynolds numbers. In particular, we train the network with the flow around airfoil data simulated between Reynolds numbers 2000 and 3000, and inspect its behavior when applied to other flow around airfoil cases within the Reynolds number range $\operatorname{Re}\in[1000,4000]$. In the remaining part of this subsection, we will first test over the stability of the predictions, and then inspect various aspects of the predictions within the stability range.

\paragraph*{Stability range} We first identify the range of Reynolds numbers within which the network is able to produce stable outputs by generating roll-out predictions for 4000 time steps for each of the test data sets (cf. Table \ref{tab:traintestsets}), starting from the first time step in each of them. The coefficient of determinations at roll-out prediction time step 4000 for different test data sets are plotted in Fig. \ref{fig:afstability}, and it is observed that the predictions within the range of Reynolds number $[2050,3200]$ are accurate and stable. This, combined with the fact that the predictions are stable for all the cross-validation data sets, leads to the conclusion that the network is able to generate stable and accurate predictions for Reynolds number range of $[2000,3200]$, i.e., the network is an effective surrogate model within the interpolation range of Reynolds numbers, and can extrapolate moderately to higher Reynolds number cases out of the training range. 

\begin{figure}
	\centering
	\includegraphics[]{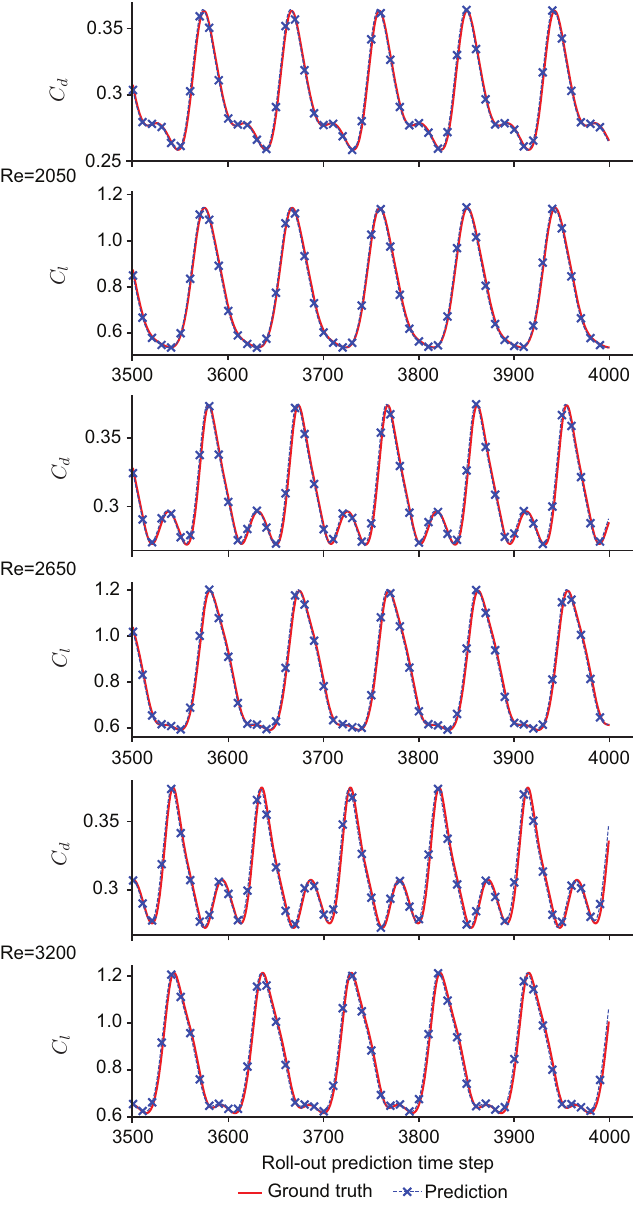}
	\caption{Predicted versus ground truth lift and drag coefficients between prediction time steps 3501 and 4000 for the flow around airfoil test data sets with Reynolds number 2050, 2650, and 3200.}
	\label{fig:liftdragaf}
\end{figure}

\paragraph*{Lift and drag coefficients} 
We plot out the lift and drag coefficients between roll-out prediction time steps 3501 and 4000 for several test data sets within the stability range in Fig. \ref{fig:liftdragaf}. It is clear that the lift and drag coefficients calculated from the flow fields predicted by the network are accurate within the stability range. One might notice the existence of slight phase differences between the predicted and ground truth lift and drag coefficient curves. This is caused by a very slight error ($\ll1\%$) in the learned vortex shedding frequencies, which is reasonable since the network is trained with one-step supervision only (cf. Sec. \ref{sec:setup}).

\begin{figure}
	\centering
	\includegraphics[]{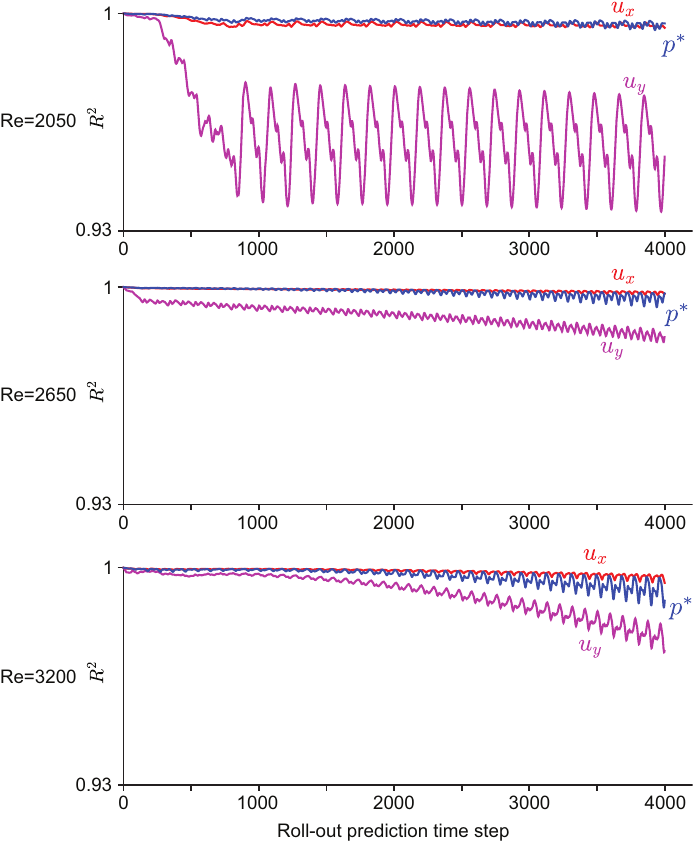}
	\caption{Coefficient of determination for the predicted velocity components $\boldsymbol{u}_x$ and $\boldsymbol{u}_y$, as well as the non-dimensionalized pressure field $\boldsymbol{p}^*$, over 4000 roll-out prediction time steps, starting from the system state at the first time step within the flow around airfoil test data sets at Reynolds number 2050, 2650, and 3200.}
	\label{fig:r2af}
\end{figure}

\paragraph*{Field predictions}
To evaluate the accuracy of the predicted systems states, we calculate the coefficient of determination $R^2$ of the predicted velocity components and pressure fields, and these values for several test data sets are plotted in Fig. \ref{fig:r2af}. The high $R^2$ values show that the predictions are stabilized and reasonably accurate. The gradual reducing trend of the $R^2$ values are caused by the accumulated phase difference over the prediction roll-out. 

\begin{figure}
	\centering
	\includegraphics[]{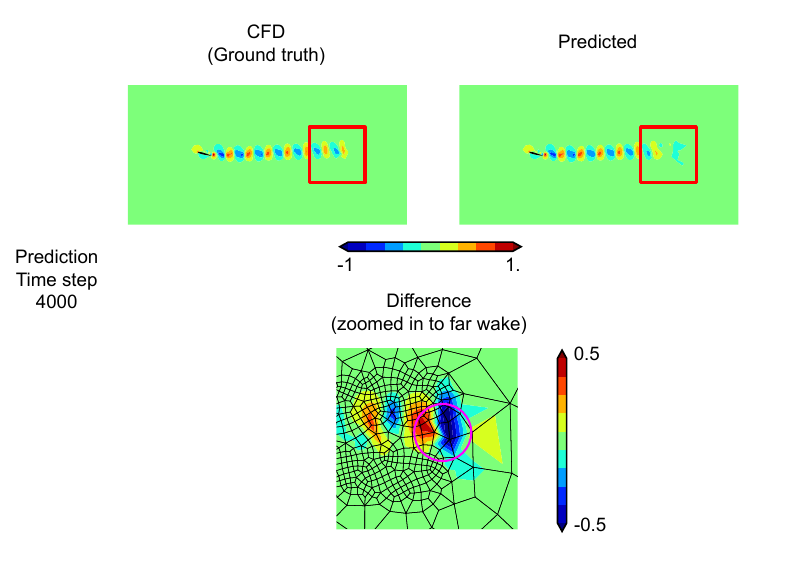}
	\caption{Predicted versus ground truth $y$-axis velocity component $\boldsymbol{u}_y$ at roll-out prediction time step 4000, for the flow around airfoil test data set at Reynolds number 2050. The coarsening of the mesh used for the neural network in this region leads to the error of the predicted velocity fluctuations at the end of the far wake, as marked out by the magenta circle.}
	\label{fig:uyaf2050full}
\end{figure}

\paragraph*{Predictions at far wake}
We notice that the coefficient of determination of $y$-axis velocity component $\boldsymbol{u}_y$, although still close to 1, is significantly lower than others, and also demonstrate relatively large fluctuation in its values. To explore the reason behind such behavior, we plot out the predicted versus ground truth $\boldsymbol{u}_y$ fields for the whole domain at prediction time step 4000 for the test data set with Reynolds number 2050 in Fig. \ref{fig:uyaf2050full}. It turns out that the network prediction in a region near the end of the far wake region deviates relatively largely from the ground truth values, as the predicted $\boldsymbol{u}_y$ dissipates. As marked out by the magenta circle in Fig. \ref{fig:uyaf2050full}, such an error is likely caused by the coarsening of the mesh in this region. The predictions for test data sets at other Reynolds numbers also demonstrate similar (but not as large) error at about the same region.

\begin{figure}
	\centering
	\includegraphics[]{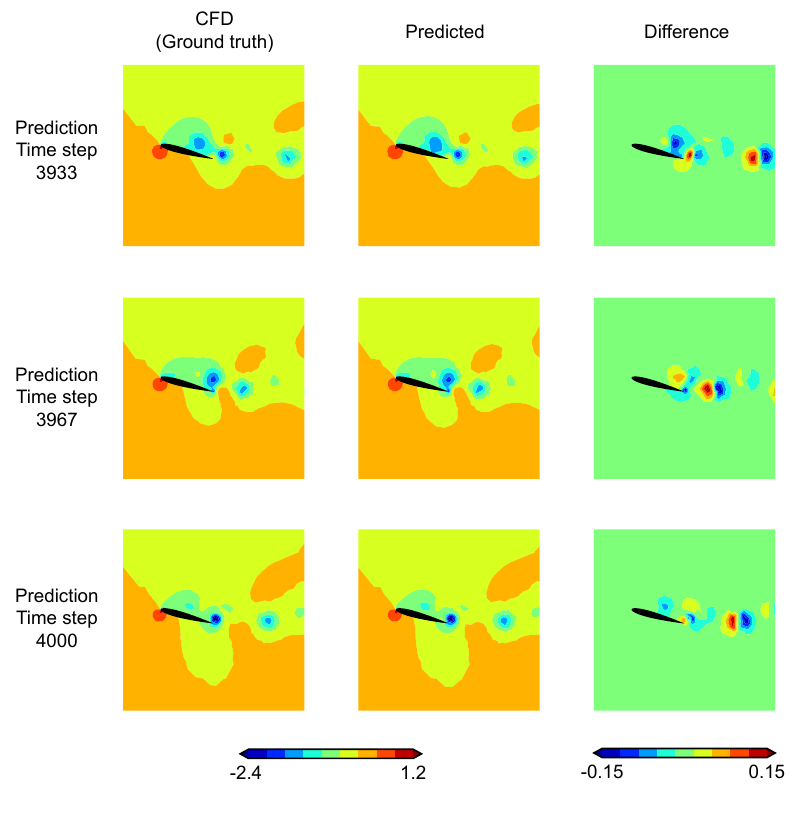}
	\caption{Predicted versus ground truth non-dimensionalized pressure field $\boldsymbol{p}^*$ at roll-out prediction time steps 3933, 3967 and 4000, zoomed in to the near wake of the airfoil, for the flow around airfoil test data set with Reynolds number 2050.}
	\label{fig:pstarafplot}
\end{figure}

\begin{figure}
	\centering
	\includegraphics[]{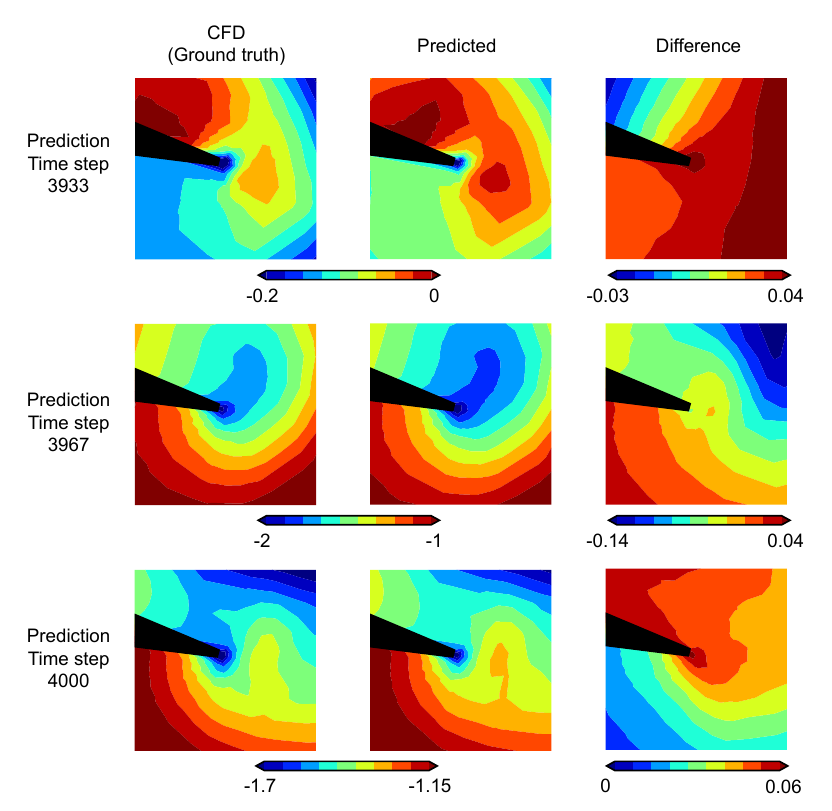}
	\caption{Predicted versus ground truth non-dimensionalized pressure field $\boldsymbol{p}^*$ at roll-out prediction time steps 3933, 3967 and 4000, zoomed in to the local neighborhood of the airfoil trailing edge, for the flow around airfoil test data set with Reynolds number 2050.}
	\label{fig:pstarafplotzoomin}
\end{figure}

\paragraph*{Predictions at near wake \& trailing edge}
We proceed to examine the predictions at near wake regions. The predicted pressure fields at time step 3933, 3967 and 4000 for the case with Reynolds number 2050 in the near wake region are plotted in Fig. \ref{fig:pstarafplot}. It is clear that the predictions are accurate in the near wake region. The prediction error is mostly due to the slight phase difference between the predicted and ground truth system states. We further zoom in to the local neighborhood of the trailing edge, and plot out the predicted versus ground truth pressure fields in Fig. \ref{fig:pstarafplotzoomin}. The network is able to predict the low pressure region at the trailing edge, as well as the flow dynamics around it.

\section{Conclusion}
\label{sec:summary}
In this paper, we presented a new graph neural network framework, and applied it to fluid dynamics simulations. Inspired by the finite element method, we converted the computational mesh to a node-element hypergraph, and designed a message-passing network on such a hypergraph that mimics the local stiffness matrix calculation process. We equipped the network with appropriate invariance and equivariance properties via input and output feature transformations. We demonstrated the proposed network for surrogate modeling of the flow around cylinder and airfoil configurations, and obtained stabilized and accurate temporal predictions for a range of Reynolds numbers. As the proposed network architecture only changes the graph connectivity and message-passing strategy at individual graph levels, it should be compatible after minor modifications with most existing techniques that were originally built upon graph neural networks defined on a normal graph. This means that the proposed network could serve as an alternative to the generalized graph message-passing network in these applications. Similar to the finite element method versus the finite volume method, we believe that some problems are more easily resolved with the proposed $\phi$-GNN framework. We thus hope that this work would inspire new applications of graph neural networks in the simulation of continuum mechanics.

It should be mentioned again that we took inspiration from the graph connectivity of the finite element method as well as the calculation process of the local stiffness matrices and load vectors. However, it remains untouched whether it is possible to construct a network architecture that is strictly a neural network version of the finite element method, which requires a step further from this work. This possibility will be further investigated in our future work.

\section*{Acknowledgment}
This research is funded by the Natural Sciences and Engineering Research
Council of Canada (NSERC) and Seaspan Shipyards. The training and evaluation of the neural network models were supported in part by the computational resources and services provided by Advanced Research Computing at the University of British Columbia. Dr. Renjie Liao and Xiaoyu Mao provided numerous suggestions to the work, their help is hereby gratefully acknowledged.

\bibliography{references}

\appendix
\appendix
\setcounter{equation}{0}
\setcounter{figure}{0}
\section{Necessity of a Reynolds number feature}
\label{sec:withre}
\begin{figure}
	\centering
	\includegraphics[]{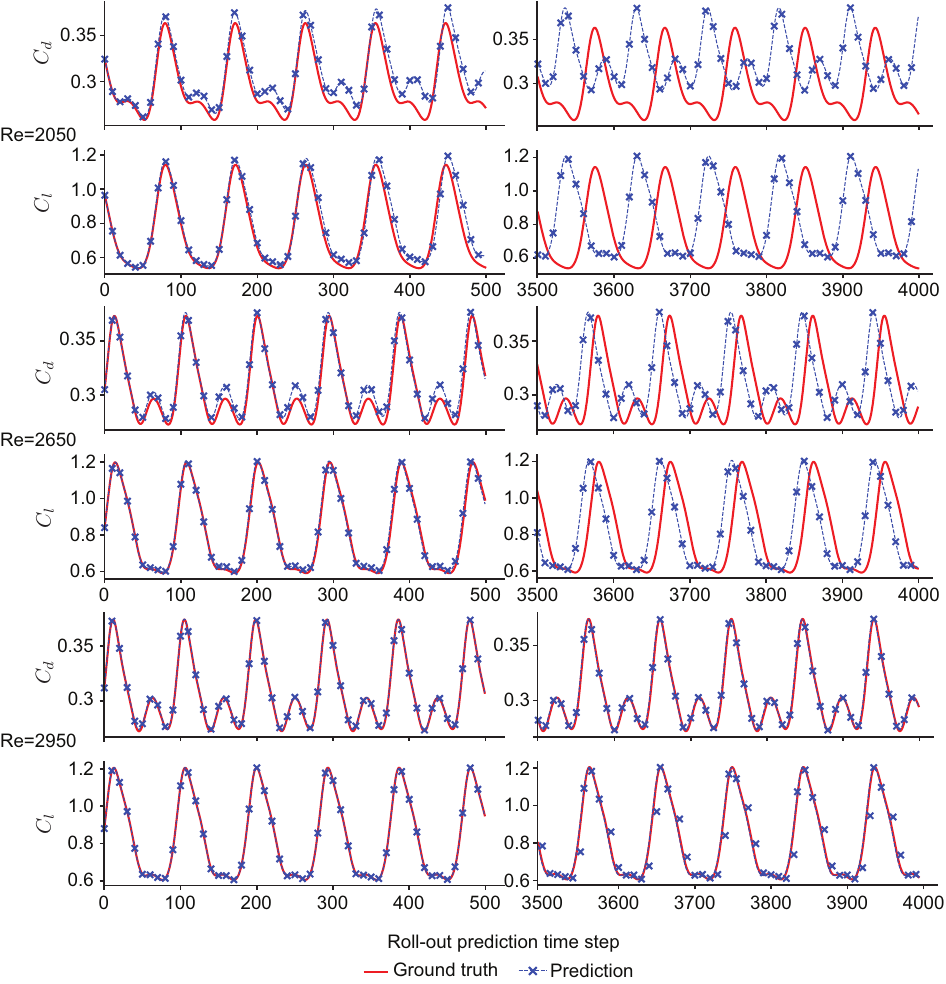}
	\caption{Predicted versus ground truth lift and drag force curves between prediction time step 1 to 500 and 3501 to 4000, for the flow around airfoil test data sets with Reynolds number 2050, 2650 and 2950, when Reynolds number is not explicitly supplied as a input feature to the network.}
	\label{fig:afnore}
\end{figure}

The Reynolds number is explicitly used as an input feature for the network in this work (cf. Eq. \ref{eq:nefeataf}). While theoretically it should be possible to implicitly identify dynamics at different Reynolds numbers, practically the network do not give good results when the Reynolds numbers are not explicitly supplied. To demonstrate this, we train the network with the airfoil data set with almost the same setup as the model reported in Sec. \ref{sec:resultsaf}, except that the element-node edge feature vectors are constructed with Eq. \ref{eq:nefeatcyl} and that noise with $3\times10^{-2}$ are added rather than $2.5\times10^{-2}$ to stabilize the outputs. The resulting lift and drag force curves between prediction time steps 1 to 500 as well as 3501 to 4000 are plotted in Fig. \ref{fig:afnore}. The network is still able to generate stable predictions, but the results are not correct. In particular, the lift and drag force curves are gradually 'corrected' towards a wrong pattern over the first hundred time steps. The similarity in shape between the predicted force curves for test sets at different Reynolds numbers indicate that the network is not able to distinguish between the dynamics at different Reynolds numbers from the flow data directly, but rather treats all cases as if they are governed by one certain dynamics. Such a behavior is consistent with the observations from our recent work \cite{deo2023combined}, where it is shown that a convolutional neural network encounter difficulties in learning the difference between the flow dynamics at different Reynolds numbers when it is trained to learn the flow evolution from a single time step to another single time step, while being able to do so when the flow information from a series of continuous time steps are used as input.

\section{Effect of training noise amount}
\label{sec:noisemag}
\begin{figure}
	\centering
	\includegraphics[]{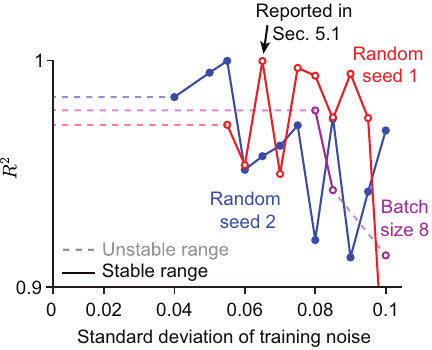}
	\caption{Coefficient of determination of the predicted pressure field $\boldsymbol{p}^*$ at prediction time step 4000 when different amounts of training noise are used. Red line: Models trained with the setup described in Sec. \ref{sec:exp} with different amounts of training noise. Blue line: Changing random seed to 2 from the setup described in Sec. \ref{sec:exp}. Magenta line: Changing batch size to 8 from the setup described in Sec. \ref{sec:exp}.}
	\label{fig:cylnoisemags}
\end{figure}

As mentioned in Sec. \ref{sec:noise}, the optimal amount of training noise is difficult to determine. To further explore the behavior of the network with different amounts of training noise, we plot out the coefficient of determination of the predicted non-dimensionalized pressure field $p^*$ at prediction time step 4000 when different amounts of training noise are used in Fig. \ref{fig:cylnoisemags}. The results show that the trend is rather chaotic, with training batch size, random seed, and the total number of training iterations influencing the optimal choice of training noise. The only conclusion that we could draw from these results is that the network would not produce stable results when not enough amount of noise is added during the training process.

It is likely that these conclusions would also apply to a wide range of other scenarios in which training noise is used, and this lead to a greater question when comparing two (or multiple) different networks: How to find a certain set of training scenarios that are completely not biased towards any of the networks in comparison, or does such a set of setup even exist? One might think about comparing the networks without using training noise, but such a comparison could be biased against those networks that are designed to work with training noise. \emph{Exactly due to this issue, we avoid comparing prediction accuracy with other existing graph neural network architectures in this work.} Nevertheless, the development of unbiased, rigorous criteria to compare different neural networks is certainly a core issue in neural network-based surrogate modeling of fluid systems which can be of interest for our future investigations.

%\end{linenumbers}

\end{document}